\documentclass[aps,prd,onecolumn,floatfix,superscriptaddress,groupedaddress,nofootinbib,preprintnumbers]{revtex4-2}

\usepackage{rotating}
\usepackage{array}
\usepackage{amsmath}
\usepackage[normalem]{ulem}
\usepackage{slashed}
\usepackage[pdftex,table]{xcolor}
\usepackage{units}
\usepackage{amssymb}
\usepackage{url}
\usepackage{comment}

\usepackage{multirow}
\usepackage{hyperref}

\newcommand\cg{\textsc{CaloGAN}}
\newcommand\cf{\textsc{CaloFlow}}
\newcommand\cfvI{\textsc{CaloFlow v1}}
\newcommand\cfvII{\textsc{CaloFlow v2}}
\newcommand\geant{\textsc{Geant}4}

\DeclareMathOperator{\MSE}{MSE}

\def\beq{\begin{equation}}
\def\eeq{\end{equation}}
\newcommand{\bea}{\begin{eqnarray}\begin{aligned}}
\newcommand{\eea}{\end{aligned}\end{eqnarray}}

\begin{document}

\title{\boldmath \cf\ II:  Even Faster and Still Accurate Generation of Calorimeter Showers with Normalizing Flows}
\author{Claudius Krause}
\email{Claudius.Krause@rutgers.edu}
\author{David Shih}
\email{shih@physics.rutgers.edu}
\affiliation{\normalsize NHETC, Dept. of Physics and Astronomy, Rutgers University, Piscataway, NJ 08854, USA}

\begin{abstract}
  Recently, we introduced \cf, a high-fidelity generative model for \geant\ calorimeter shower emulation based on normalizing flows. Here, we present \cfvII, an improvement on our original framework that  speeds up shower generation by a further factor of 500 relative to the original. The improvement is based on a technique called Probability Density Distillation, originally developed for speech synthesis in the ML literature, and which we develop further by introducing a set of powerful new loss terms. We demonstrate that \cfvII\ preserves the same high fidelity of the original using qualitative (average images, histograms of high level features) and quantitative (classifier metric between \geant\ and generated samples) measures. The result is a generative model for calorimeter showers that matches the state-of-the-art in speed (a factor of $10^4$ faster than \geant) and  greatly surpasses the previous state-of-the-art in fidelity.
\end{abstract}

\maketitle
\flushbottom

\section{Introduction}

The enormously successful physics program at the LHC relies heavily on the availability of copious amounts of highly accurate simulated data. However, the use of \geant~\cite{Agostinelli:2002hh,1610988,ALLISON2016186} for full detector simulations is a major computational bottleneck and severely limits the analysis capabilities of the LHC. This is forecast to worsen significantly with future LHC upgrades and the HL-LHC~\cite{Apostolakis:2018ieg,Aarrestad:2020ngo,Calafiura:2729668,CMS:computing,ATLAS:2021pzo}. 

Recently, deep generative modeling  has demonstrated great potential to speed up the most computationally expensive part of detector simulations, namely calorimeter showers \cite{Paganini:2017hrr,Paganini:2017dwg,deOliveira:2017rwa,Erdmann:2018kuh,Erdmann:2018jxd,ATL-SOFT-PUB-2018-001,Belayneh:2019vyx,Buhmann:2020pmy,Buhmann:2021lxj,ATL-SOFT-PUB-2020-006, Krause:2021ilc, ATLAS:2021pzo}. By fitting the generative model to \geant\  shower images, the generative model learns (often implicitly) the underlying distribution that the \geant\ showers are drawn from and can then sample from it quickly. Most of the current approaches \cite{Paganini:2017hrr,Paganini:2017dwg,deOliveira:2017rwa,Erdmann:2018kuh,Erdmann:2018jxd,ATL-SOFT-PUB-2018-001,Belayneh:2019vyx,Buhmann:2020pmy,Buhmann:2021lxj,ATL-SOFT-PUB-2020-006, ATLAS:2021pzo} are based on GAN or VAE architectures. Very recently, in~\cite{Krause:2021ilc}, we proposed a fresh alternative, dubbed \cf, based on normalizing flows (for recent reviews and original references, see e.g.~\cite{2019arXiv190809257K,2019arXiv191202762P}). Flows have many advantages over GANs and VAEs --- stable, convergent, principled training and model selection, based on minimizing the negative log likelihood (NLL) objective; no issues with mode collapse; explicitly learning a differentiable likelihood function; and fully bijective mapping to latent space. 
Correspondingly, we found that the quality of the generated samples was astonishing; unlike the baseline GAN that we compared with \cite{Paganini:2017hrr,Paganini:2017dwg}, the samples produced with \cf\ were almost indistinguishable from \geant\ samples~\cite{Krause:2021ilc}.\footnote{It is possible that the fidelity of the GAN could be improved with more sophisticated architectures compared to the original CaloGAN \cite{Paganini:2017hrr,Paganini:2017dwg}; however, see \cite{Diefenbacher:2020rna} for an example of a state-of-the-art GAN architecture that also produces samples that can be easily distinguished from the training data with a classifier.} 

However,  the approach presented in~\cite{Krause:2021ilc}  had one major downside compared to the other deep generative models: sampling speed. While \cf\ was $\sim50$ times faster than \geant, it was considerably slower (by a factor of  $\sim500$) than the GAN-based alternative. The disadvantage in generation speed was because of the Masked Autoregressive Flow (MAF) architecture~\cite{2017arXiv170507057P} used in~\cite{Krause:2021ilc}, which is only fast in density estimation, but a factor 500 (given by the dimensionality of the read-out channels of the detector) slower in sampling. Normalizing flows can also be constructed to be fast in the other direction (fast sampling, slow density estimation); this goes by the name of Inverse Autoregressive Flow (IAF)~\cite{2016arXiv160604934K}. Indeed, using an IAF would bring \cf\ fully in line with the speed of the GAN-based approaches.
However, what sounds like the solution does not work in practice, as IAFs with such a high dimensionality cannot be trained using the NLL objective, due to time and memory limitations, as we will elaborate on. 

In this paper, we overcome the challenges of training the IAF, by building on a method alternately called {\it Probability Density Distillation} or {\it teacher-student training} in the ML literature. Originally developed for the purposes of speech synthesis in~\cite{2017arXiv171110433V}, the basic idea is that the IAF (called the \emph{student}) can be trained efficiently not on the original target data, but on the output of a trained MAF model (called the \emph{teacher}). The MAF can quickly map target data points to the latent space, $x\to z$, and the IAF inverse can quickly map the latent space back to the target data space, $z\to x'$. By requiring this loop to close ($x'=x$), i.e.\ by requiring the MAF and IAF to describe the same transformation, or, in other words, the fast passes to be each others inverses, one can in principle get an arbitrarily good fit of the IAF to the MAF (and by extension, to the original target data).

The original idea of Probability Density Distillation presented in~\cite{2017arXiv171110433V} relied on minimizing the KL divergence between the IAF and MAF. However, even in that work, they found that  this training objective was insufficient, and they added additional, ad hoc loss terms based on high-level features in order to get the IAF to converge to the MAF.  Subsequently, \cite{pmlr-v115-huang20c} explored more well-motivated and general alternatives to the KL divergence loss, based on the distance between $x$ and $x'$ in the MAF-IAF loop, or analogously $z$ and $z'$ in the IAF-MAF loop. These provide an alternative measure of closure that has improved convergence and convexity properties compared to the KL divergence. Other works in the ML literature that explore variants on the idea of Probability Density Distillation include \cite{2021arXiv210612699B,yamamoto19b_interspeech,2018arXiv180707281P}.

Here we build on the version of Probability Density Distillation presented in  \cite{pmlr-v115-huang20c}. We found that while using $x$--$x'$ and $z$--$z'$ distances offered significant improvements to the KL divergence, still they were insufficient for achieving a good fit of the IAF to the MAF. But by adding additional measures of the MAF-IAF closure, involving the intermediate layers and actual transformation parameters of the invertible normalizing flow (see sec.~\ref{sec:PDD} for details), we were able to obtain an excellent fit of the IAF to the MAF. Along the way, we also devise a new method for model selection of the IAF. Since the NLL is too expensive to compute for every epoch, we instead use the much cheaper KL divergence. Although the KL divergence is not a suitable objective for training the IAF, we show that it nevertheless tracks the NLL very closely, and therefore can serve as an effective proxy for selecting the best model epoch.

The result is a new version of \cf\ (and a new version of Probability Density Distillation!) that is just as fast as the GAN baseline, while just as high-fidelity as the MAF used in \cite{Krause:2021ilc}. We demonstrate this using the same qualitative measures as in \cite{Krause:2021ilc}, as well as the classifier metric introduced in \cite{Krause:2021ilc}. 

The outline of our paper is as follows. Section~\ref{sec:MAFIAF} reviews the construction of the MAF and IAF normalizing flows, and explains why the former is fast to density estimate and slow to sample, while the latter is the opposite. Section~\ref{sec:PDD} explains the idea behind Probability Density Distillation and describes our new loss terms that greatly improve the matching of the student to the teacher. Section~\ref{sec:calo} very briefly describes the dataset used for this work; for more details see  \cite{Krause:2021ilc,Paganini:2017hrr,Paganini:2017dwg}. Section~\ref{sec:cf} describes the architecture and training procedure of \cfvII.
Section~\ref{sec:results} contains the results of the teacher-student training --- average images, histograms, classifier metric and timing. Finally we summarize and conclude in section~\ref{sec:conclusions}. In Appendix~\ref{app:nearest}, we present plots of nearest neighbors between \geant\ and \cf\ student samples, providing further evidence against mode collapse in the latter.

\section{Density Estimation and Probability Distillation with Normalizing Flows}
\label{sec:DE}

\subsection{MAFs vs.\ IAFs: fast and slow directions}
\label{sec:MAFIAF}

Our work uses autoregressive flows to learn the invertible transformation between the data space $x\in {\Bbb R}^d$ and the latent space $z\in {\Bbb R}^d$. These autoregressive flows take the form 
\beq
z_i = f(x_i;\vec{\kappa}_i), \qquad x_i=f^{-1}(z_i;\vec{\kappa}_i),\qquad i=1,\dots,d
\eeq
where $f$ is an invertible 1d transformation (we use Rational Quadratic Splines~\cite{NEURIPS2019_7ac71d43, 10.1093/imanum/2.2.123}), and $\vec{\kappa}_i$ are a set of coordinate-dependent parameters for the transformation of the $i$-th coordinate. To preserve the autoregressive property, the parameters should only depend on the previous coordinates (i.e.\ those with index less than $i$). But we have a choice as to whether we make the parameters explicit functions of the $x$ coordinates or the $z$ coordinates. That is, we must have either
\beq\label{eq:pofx}
\vec{\kappa}_i=\vec{\kappa}_i(x_1,\dots,x_{i-1}) 
\eeq
or
\beq\label{eq:pofz}
\vec{\kappa}_i=\vec{\kappa}_i(z_1,\dots,z_{i-1})
\eeq
In our setup, all these parameters are the output of a neural network (the MADE block, from ``Masked Autoencoder for Distribution Estimation''~\cite{2015arXiv150203509G}), and the autoregressive property is accomplished through the application of a binary mask on the internal hidden layers. 

The first choice, eq.~\eqref{eq:pofx}, defines the Masked Autoregressive Flow (MAF) architecture~\cite{2017arXiv170507057P}. When the parameters are explicit functions of $x_i$, then $x\to z$ (the ``forward'' pass for ``inference'' aka density estimation) is fast, requiring just a single evaluation of the neural networks. However, to perform the inverse transformation $z\to x$ (for sampling), one must compute
\beq
x_i=f^{-1}(z_i,\vec{\kappa}_i(x_1,\dots,x_{i-1}))
\eeq
Now the $x_1,\dots,x_{d}$ are only known recursively, i.e.\
\bea
\label{eq:inverse}
x_1 &=f^{-1}(z_1,\vec{\kappa}_1)\\
 x_2 &=f^{-1}(z_2,\vec{\kappa}_2(x_1))=f^{-1}(z_2,\vec{\kappa}_2(f^{-1}(z_1,\vec{\kappa}_1)))\\
 x_3 &=f^{-1}(z_3,\vec{\kappa}_3(x_1,x_2))=f^{-1}(z_3,\vec{\kappa}_3(f^{-1}(z_1,\vec{\kappa}_1),f^{-1}(z_2,\vec{\kappa}_2(f^{-1}(z_1,\vec{\kappa}_1)))))\\
 &\dots
\eea
So to evaluate the inverse transformation $z\to x$ requires $d$ successive evaluations of the neural network. The MAF is fast to density estimate but a factor of $d$ slower to sample.

The second choice, eq.~\eqref{eq:pofz}, defines the Inverse Autoregressive Flow (IAF) architecture~\cite{2016arXiv160604934K}. 
In that case the opposite is true: sampling $z\to x$ is fast, while density estimating $x\to z$ is a factor of $d$ slower\footnote{Note that the ``forward'' direction of the IAF also refers to density estimation.}. Unfortunately, this also means that training an IAF with the LL objective (which requires $x\to z$ density estimation) would take a factor $d$ more time than training the MAF.\footnote{In addition, since each coordinate transformation depends on the result of a pass through the network and the previous coordinates, which in turn depend on passes through the network (see eq. \eqref{eq:inverse}), the memory needed to store the gradients exceeds the memory requirement of a MAF by a large factor. In principle, one could be able to optimize the storage of these gradients similar to backpropagation, but this is currently not implemented in any of the available ML code frameworks.} Given that the MAF takes approximately $\sim{\mathcal O}(1~{\rm hr})$ to train, and $d\sim 500$ in our setup, 
 this means that the IAF would be prohibitively time consuming for us to train. Instead, training an IAF generative model for calorimeter showers requires a very different approach. 

\subsection{Probability density distillation and teacher-student training}
\label{sec:PDD}

The key idea for how to train an IAF efficiently, which we build upon in this work, was introduced in~\cite{2017arXiv171110433V} in the context of speech synthesis, and given the name of \emph{Probability Density Distillation} or \emph{teacher-student training}. The idea is that while fitting the IAF directly to data is practically prohibitive, fitting the IAF to the MAF is not. By starting from a sample $z$ in the latent space and mapping it to data space via  the {\it student} (IAF), we get a set of data samples $x$ and their likelihood $s(x)$ under the student efficiently. Mapping it back to latent space with the {\it teacher}  (MAF) yields the log-likelihood of the same sample under the teacher, $t(x)$. Every pass is then the fast one under its respective autoregressive flow.

In~\cite{2017arXiv171110433V}, the loss function was initially taken to be the KL divergence between these two probability densities, 
\begin{equation}
  \label{eq:KL}
  \text{KL} = \int s(x) \log{\frac{s(x)}{t(x)}}~dx = \sum_{x\sim S} \log{\frac{s(x)}{t(x)}}.
\end{equation}
Note that this KL divergence is based on the same $x$. Starting from data and closing the loop through the teacher first and then through the student gives different $x$ and $x^{\prime}$ at the beginning and the end of the chain, so instead one would need to calculate KL in $z$ space which is not meaningful, since the base distributions of teacher and student are identical. 

Although in principle the KL divergence of eq.~\eqref{eq:KL} is a good loss --- it is non-negative and zero iff the IAF and MAF densities agree --- it was already found in~\cite{2017arXiv171110433V} to not converge well to the desired result.  The authors of \cite{2017arXiv171110433V} added additional, ad hoc high-level-feature-based loss terms to enhance the quality of their generated audio sample.

Reasons for why the KL divergence has poor convergence properties as a loss function were given in  \cite{pmlr-v115-huang20c}. Since the IAF fast pass should be the inverse of the MAF fast pass, two alternative loss functions were proposed:
\beq\label{eq:Lx}
L_{x} \equiv \MSE(x, x^{\prime})  
\eeq
and
\beq\label{eq:Lz}
L_{z} \equiv \MSE(z, z^{\prime})
\eeq
Here, one can start from data $x$, map it to latent space $z$ with the MAF, and map it back to $x'$ in the data space with the IAF; every pass is then the fast one under its respective autoregressive flow. Starting from noise and mapping back to noise, $z\to x\to z'$ via IAF and then MAF is also possible. Either way, requiring that the transformation closes forces the IAF to conform to the MAF. 
These measure more directly the closure of the MAF-IAF loop.\footnote{In principle any distance measure in the data space and latent space could be used; for simplicity we used simple Euclidean distance in image space and in the latent space and empirically this worked well.}

In this work, we go beyond the loss functions of eqs.~\eqref{eq:Lx} and~\eqref{eq:Lz}, as we observed that while they do improve on the KL divergence (which generally doesn't converge at all), they still lead to poor overall agreement between the IAF and the MAF and a bad NLL of the trained student flow.  

\begin{itemize}

\item First, we found that using both $L_x$ and $L_z$ together (we tried the simple average of the two) was much better than using each one separately as was  considered in \cite{pmlr-v115-huang20c}.

\item The MAF and IAF actually parametrize a series of invertible autoregressive transformations. Looking at the fast passes through the flows, we can think of them as 
\beq
x\to y^{(1)}_{t}\to y^{(2)}_{t}\to\dots\to y^{(N)}_{t} = z\qquad \text{(MAF)}
\eeq
and
\beq
z\to y^{(1)}_{s}\to y^{(2)}_{s}\to\dots\to y^{(N)}_{s} = x\qquad \text{(IAF inverse)}
\eeq
Here, $y_{t}^{(i)}$ is understood to include the permutation after the transformation and $y_{s}^{(i)}$ the permutation before the transformation\footnote{The permutation between the latent space and the adjacent MADE block is absorbed in the permutation-invariant base distribution.}. Since each step is an invertible transformation, we could require that the MAF and IAF to agree with each other at every step, and not just at the endpoints. In other words, we could require that
\beq
y^{(1)}_{s}=y^{(N-1)}_{t},\quad y^{(2)}_{s}=y^{(N-2)}_{t},\quad\dots\quad,\quad y^{(N-1)}_{s}=y^{(1)}_{t}
\eeq
If we start from $x$ the full chain of transformations in the loop is:
\beq
x\to y^{(1)}_{t}\to\dots\to y^{(N)}_{t} = z\to y^{(1)}_{s} \to y^{(2)}_{s}\to\dots\to y^{(N)}_{s} = x^{\prime}
\eeq
and we could enforce stepwise agreement with the losses
\beq\label{eq:Lxi}
L_{x^{(i)}}=\MSE(y^{(i)}_{t}(x),y^{(N-i)}_{s}(x))\,\, ,\qquad i=1,\dots,N-1
\eeq
where we have explicitly indicated here that the $y^{(i)}_{t}$ and $y^{(j)}_{s}$ are functions of (originated from) $x$.
Similarly we can perform the loop starting with $z$, and enforce stepwise agreement with
\beq\label{eq:Lzi}
L_{z^{(i)}}=\MSE(y^{(i)}_{s}(z),y^{(N-i)}_{t}(z))\,\, ,\qquad i=1,\dots,N-1
\eeq  
Including the sum of these losses in the training (in addition to $L_x$ and $L_z$) improved the student-teacher matching further. 

\item Finally, we could further exploit the constraint that the MAF and IAF parametrize the same transformation at each step, and require that the parameters output by each MADE block agree between the IAF and MAF:
\beq\label{eq:Lpxi}
L_{\kappa_x^{(i)}}=\MSE(\kappa_t^{(i)}(x),\kappa_s^{(N-i+1)}(x))\,\, ,\qquad i=1,\dots,N
\eeq
and
\beq\label{eq:Lpzi}
L_{\kappa_z^{(i)}}=\MSE(\kappa_s^{(i)}(z),\kappa_t^{(N-i+1)}(z))\,\, ,\qquad i=1,\dots,N
\eeq  
where the former (latter) set of $\kappa_{i}$ is understood to be coming from a pass that started with data $x$ (noise $z$). Since this MSE captures the full spline and not just the bin the coordinate falls into, this has the potential to drive the student even closer to the teacher than the MSEs based on eqs.~\eqref{eq:Lxi}--\eqref{eq:Lzi}. Indeed, we found that including the sum of these in the loss (together with those above) led to the best result.
\end{itemize}

\begin{table}[!t]
  \caption{NLL (smaller is better) after training \cf\ for 150 epochs on $e^{+}$ showers with different loss functions (sums over $i$ are understood for all applicable cases). As a comparison, the teacher model has a NLL of $142.2 $.} 
\label{tab:loss}
\begin{center}
  \begin{tabular}{|l|r|}
    \hline
    Loss & NLL \\
    \hline
    $L_{x}$ & $1596.3 $ \\
    $L_{z}$ & $256.6 $ \\
    $ L_{x} + L_{z}$ & $198.7 $  \\
    $L_{x} + L_{z}+ L_{x^{(i)}} + L_{z^{(i)}} $ & $170.6 $ \\
    $L_{x} + L_{z} + L_{\kappa_{x}^{(i)}}+ L_{\kappa_{z}^{(i)}}$ & $147.4 $ \\
    $L_{x} + L_{z} +  L_{x^{(i)}} + L_{z^{(i)}}+L_{\kappa_{x}^{(i)}}+ L_{\kappa_{z}^{(i)}} $ (Eq.~\eqref{eq:final.loss})  & $146.4$ \\
    \hline
  \end{tabular}
\end{center}
\end{table}

\begin{figure}[t!]
    \centering
    \includegraphics[width=\textwidth]{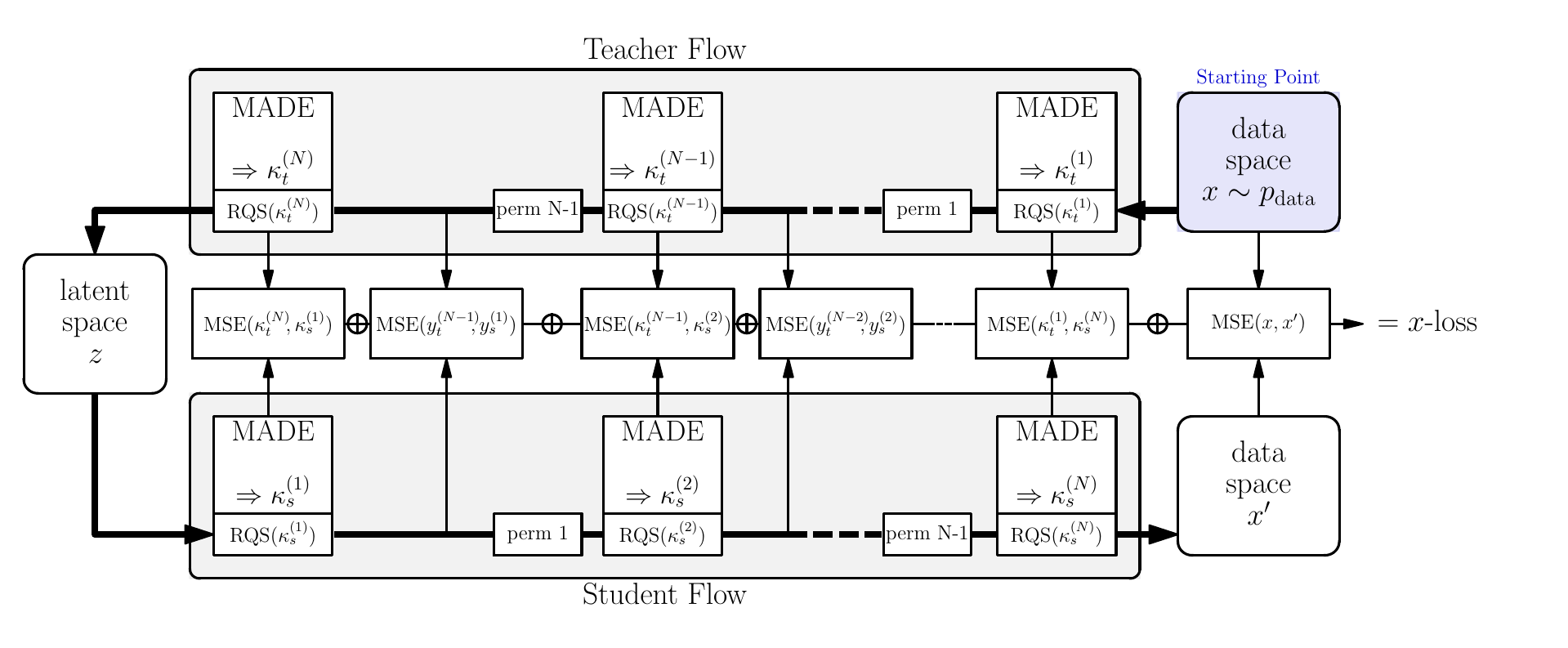}\\
    \includegraphics[width=\textwidth]{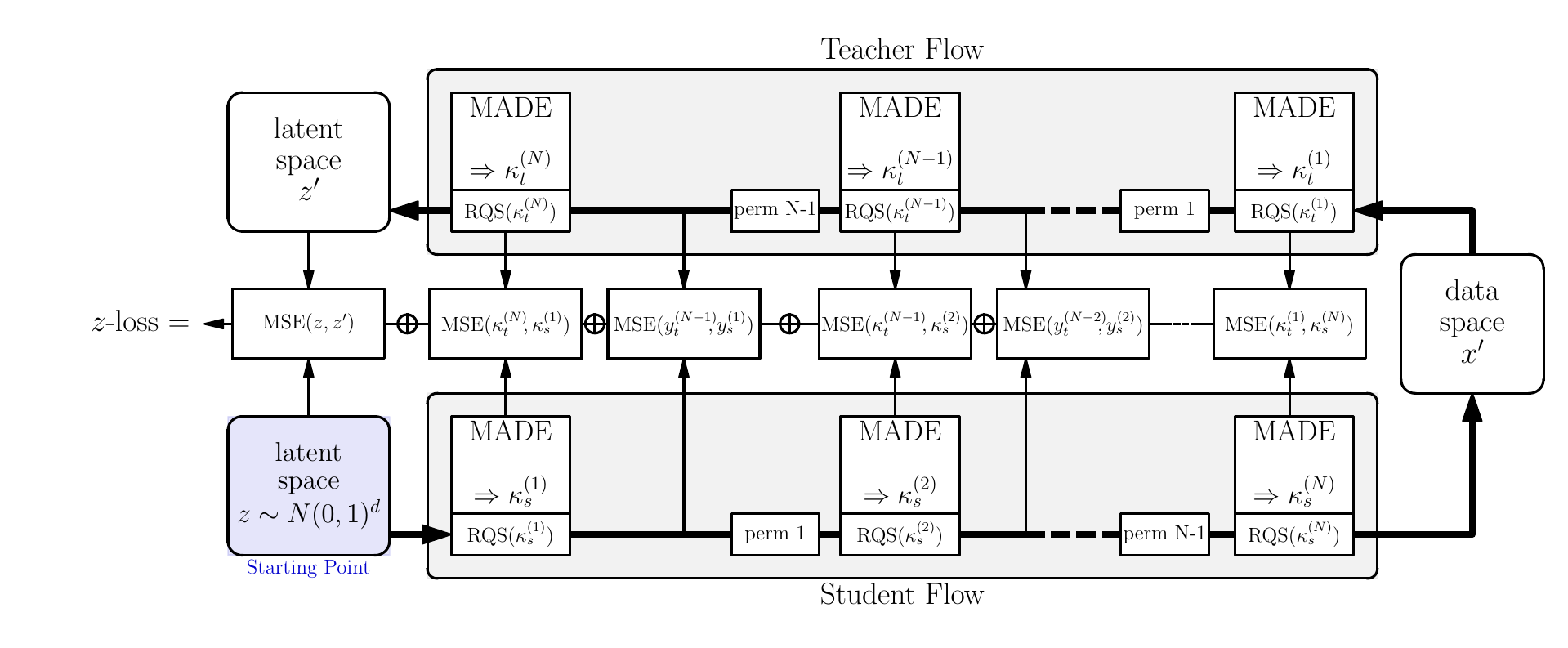}
    \caption{Schematic view of the construction of the loss function. Top: $x$-loss, actual data $x$ is fed through the teacher and student starting from the top right (indicated in blue). Bottom: $z$-loss, generated noise $z$ is fed through the student and teacher starting from the lower left (indicated in blue). Intermediate coordinates (top: $y_{t}^{(i)}$; bottom: $y_{s}^{(i)}$) and RQS parameters (top: $\kappa_{t}^{(i)}$; bottom: $\kappa_{s}^{(i)}$) are compared in an MSE loss and then summed.}
    \label{fig:loss}
  \end{figure}

The various loss terms are illustrated in fig.~\ref{fig:loss}.
In table~\ref{tab:loss}, we demonstrate the successive improvements to the NLL for $e^{+}$ showers due to including these loss terms, after training for 150 epochs  as described in section~\ref{sec:training}. We observe a clear improvement of the NLL the more terms are added to the loss. 
%\footnote{However, the difference between the last two is quite small, so not adding $L_{x^{(i)}} $ and $ L_{z^{(i)}}$ to the loss function provides a further viable loss candidate.} 
Evidently, the student does best if it is guided as closely as possible. \footnote{In principle, the student does not have to be an IAF, it could also be a simple, fully-connected neural network~\cite{2021arXiv210612699B}. However, in this case we would not have access to the LL as a measure of quality and we would not be able to train it with the additional loss terms of eqs.~\eqref{eq:Lxi}--\eqref{eq:Lpzi}.} 

In summary, our final objective function for the teacher-student training is:
\begin{align}
  \begin{aligned}
    \label{eq:final.loss}
    L &= 0.5 \left(\underbrace{\MSE(z, z^{\prime}) + \sum_{i}^{N-1}\MSE(y^{(i)}_{s}(z),y^{(N-i)}_{t}(z))+\sum_{i}^{N}\MSE(\kappa_s^{(i)}(z),\kappa_t^{(N-i+1)}(z))}_{z\text{-loss}}\right)\\
    &+ 0.5 \left(\underbrace{\MSE(x, x^{\prime}) + \sum_{i}^{N-1}\MSE(y^{(i)}_{t}(x),y^{(N-i)}_{s}(x))+\sum_{i}^{N}\MSE(\kappa_t^{(i)}(x),\kappa_s^{(N-i+1)}(x))}_{x\text{-loss}}\right)
  \end{aligned} %           
\end{align}
 We will refer to training with the objective given by eq.~\eqref{eq:final.loss} as ``fully-guided'' student training.

\section{Calorimeter Data}
  \label{sec:calo}
  
Since this is an improvement of \cfvI~\cite{Krause:2021ilc}, we use the same calorimeter setup as there, which was based on \cg\ \cite{Paganini:2017hrr,Paganini:2017dwg}. Here we provide a very brief description; we refer the reader to \cite{Krause:2021ilc} for details. The calorimeter is a simplified version of the ATLAS electromagnetic calorimeter. It has three layers of sizes $3\times 96$, $12\times 12$ and  $12\times 6$ voxels respectively. The training data are showers of $e^+$, $\gamma$ and $\pi^{+}$ with energies uniformly sampled from 1--100~GeV and perpendicularly incident on the calorimeter simulated with \geant. These are the exact same samples~\cite{caloflowdata} that were used to train and evaluate \cfvI. For each particle, we have: a set of 70,000 showers to train the flow; a set of 30,000 showers for model selection and validation of the flow; as well as additional sets of 60,000/20,000/20,000 showers to train, validate/calibrate and test the classifier metric of section~\ref{sec:cls}.

\section{\cfvII}
\label{sec:cf}

\subsection{Architecture}

As in \cite{Krause:2021ilc}, we preserve the two-step structure of \cf. In the first step, we use a small normalizing flow, called Flow I, to learn the distribution of deposited energies conditioned on the input energy, $p_1(E_{0}, E_{1}, E_{2}|E_{\mathrm{inc}})$. In the second step, we use a much larger flow, called Flow II, to learn the shower shapes conditioned on the energies, $p_{2}(\vec{\mathcal{I}}|E_{0}, E_{1}, E_{2}, E_{\mathrm{inc}})$.

Flow I is exactly as it was in~\cite{Krause:2021ilc}. In fact, we use the saved weights of~\cite{Krause:2021ilc} throughout this paper. We did not bother to train an IAF for Flow I, since the time to sample from Flow I is significantly smaller than the time to sample from Flow II, so a factor of $\sim 3$ speed-up of Flow I would have a negligible effect on the overall sampling time of \cfvII.

Instead, we focus our attention in this work on training a student IAF for Flow II, based on the teacher MAF for Flow II from \cite{Krause:2021ilc}. We use the same hyperparameters (8 blocks, 378 hidden neurons, 8 RQS bins) for the teacher as we used in~\cite{Krause:2021ilc}. (In fact, we use the saved weights of the MAF trainings from~\cite{Krause:2021ilc} for the training the student here.)  Since the student is much faster to evaluate, we could in principle make it bigger than the teacher. However, the ``fully-guided'' block-wise loss that we introduced in section \ref{sec:PDD} requires the same number\footnote{One could consider adding MADE blocks to the student and match groups of them to a single teacher MADE block, but we did not pursue this strategy here.} of MADE blocks between the teacher and student. We are left with making the hidden layers wider, and we chose 504 nodes (to match the dimensionality of the voxel space), as we found that this improved the performance of the student. Another modification we considered was to make the NN inside the MADE blocks deeper, but an initial study showed no improvement from this. Finally, we also considered making the teacher bigger (and hence slower) than in~\cite{Krause:2021ilc}; this showed no improvement in the LL of the teacher, probably due to overfitting. The IAF permutation $i$ is taken to be the same as the MAF permutation $N-i$.

\subsection{Training}
\label{sec:training}
We train the student of Flow II as follows. We use the same training and validation datasets as for the teacher in \cfvI, as we saw no sign of overfitting in this setup. For every epoch, we shuffle and  divide up the 70,000 samples of the target \geant\ training data into minibatches of 175 events. We feed these minibatches through the teacher MAF and back through the student IAF and obtain the $x$-loss and gradients with respect to the student weights. During each minibatch, we also sample 175 events from the latent space. We feed these through the student IAF and back through the teacher MAF and obtain the $z$-loss and gradients wrt student weights.  These are finally all combined together, and total loss is minimized wrt the student weights via the \textsc{Adam}~\cite{kingma2014adam} optimizer for 150 epochs. 

We found that increasing the minibatch size in training improves the convergence. To overcome memory constraints with too large minibatch sizes, we train the student using the gradient accumulation technique: the gradients of several minibatches are stored before a parameter update step is performed, effectively increasing the minibatch size. 

We also found that optimizing with a rather sophisticated learning rate schedule helped the student converge better to the teacher. We start with a learning rate of $2.5\cdot 10^{-3}$ and a minibatch size of 175. In the first epochs, the gradients of two such minibatches are accumulated before the weight update is performed. After epochs 10, 40, and 70, we apply a factor of $0.5$ to the learning rate and at the same time multiply the number of accumulated minibatches by 2. For example, after epoch 70, we accumulate 16 minibatches before the gradient update.

\begin{figure}[!th]
    \centering
    \includegraphics[width=0.95\textwidth, trim= 85 0 70 0, clip]{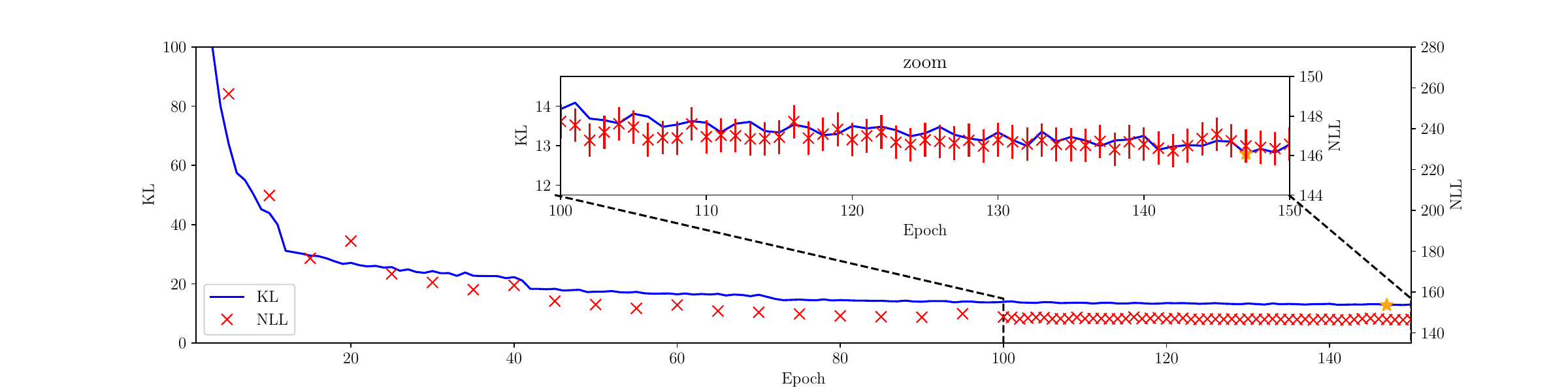}
    \caption{KL divergence and NLL during training of the student IAF for $e^+$ showers. The KL divergence is computed using 70,000 noise samples, while the NLL is computed using 10,000 \geant\ samples from the validation set. Error bars on the latter show the standard error of the NLL estimate, with the error of the KL divergence being at or below its line width. The orange star marks the epoch with smallest KL divergence.}
    \label{fig:KLvsNLL}
\end{figure}

Finally, model selection for the Flow II student is less straightforward than for the teacher. Again, since the NLL on the \geant\ validation set is very expensive to compute for the student, we cannot use this to directly select the best model. However, as described in sec.~\ref{sec:DE}, the KL divergence eq.~\eqref{eq:KL} between the teacher and student densities is very efficient to compute. Even though the KL divergence does not constitute a good loss term due to problems with its gradient \cite{pmlr-v115-huang20c}, it is still a useful metric to judge the convergence between student and teacher. In particular, we observe a strong correlation between the KL divergence and the NLL. This is illustrated in  fig.~\ref{fig:KLvsNLL} for the training of $e^{+}$-showers. 
We therefore compute the KL divergence of every training epoch and select the model state with the smallest KL divergence for the subsequent sampling and evaluation of the flow.\footnote{A computationally more expensive approach would be to save the model state based on the 5 or 10 smallest KL values and evaluate the NLL of these model states after the training. Since the weights will not be updated anymore, this evaluation could be run in parallel on different machines.}  

\section{Results}
\label{sec:results}

\subsection{Log likelihood}

We start with a comparison of total NLL between the student IAF and the teacher MAF. They are evaluated on the same validation set, containing 30,000 events drawn from the \geant\ simulation. The results are shown in table~\ref{tab:LLs} (also compare to fig.~\ref{fig:KLvsNLL}) and demonstrate that the student NLL nearly saturates the teacher NLL for all three particle types.

\begin{table}[!ht]
\caption{NLLs of the teacher and student flows, evaluated on the same validation set (lower is better). } 
\label{tab:LLs}
\begin{center}
  \begin{tabular}{|c|c|c|}
    \hline
    Particle & \cfvI\ (teacher) NLL from~\cite{Krause:2021ilc} & \cfvII\ (student) NLL\\
    \hline
    $e^{+}$ & $142.159 $ & $146.393$ \\
    $\gamma$ & $194.064 $ & $197.347$ \\
    $\pi^{+}$ & $637.265 $ & $639.678$ \\
    \hline
  \end{tabular}
\end{center}
\end{table}

\subsection{Average images}

\begin{figure}[!ht]
    \centering

    \includegraphics[width=0.3\textwidth]{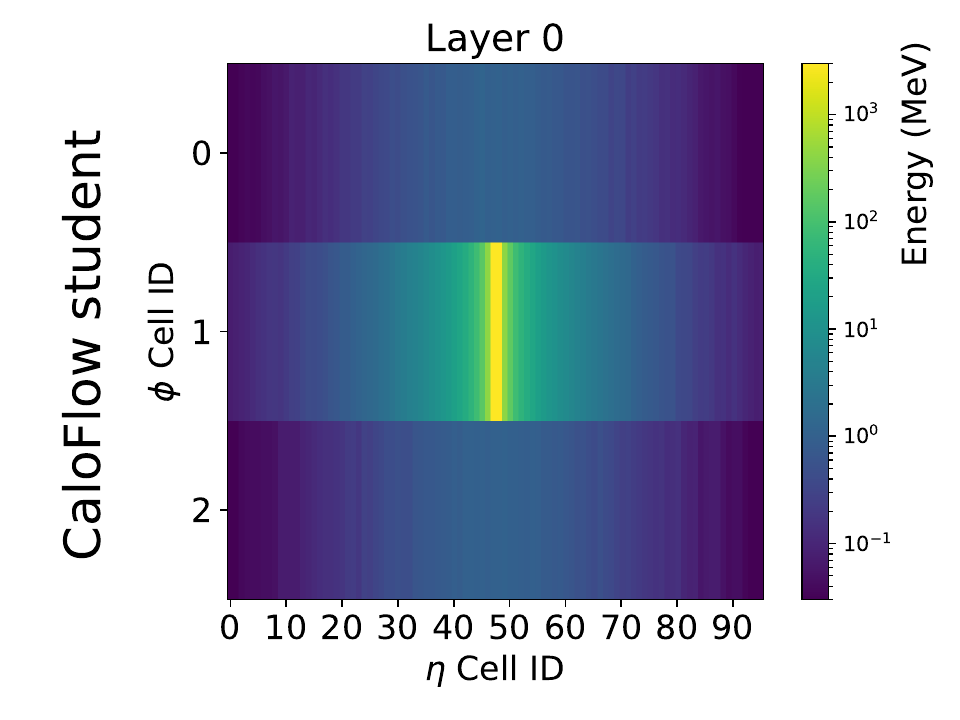}
    \includegraphics[width=0.3\textwidth]{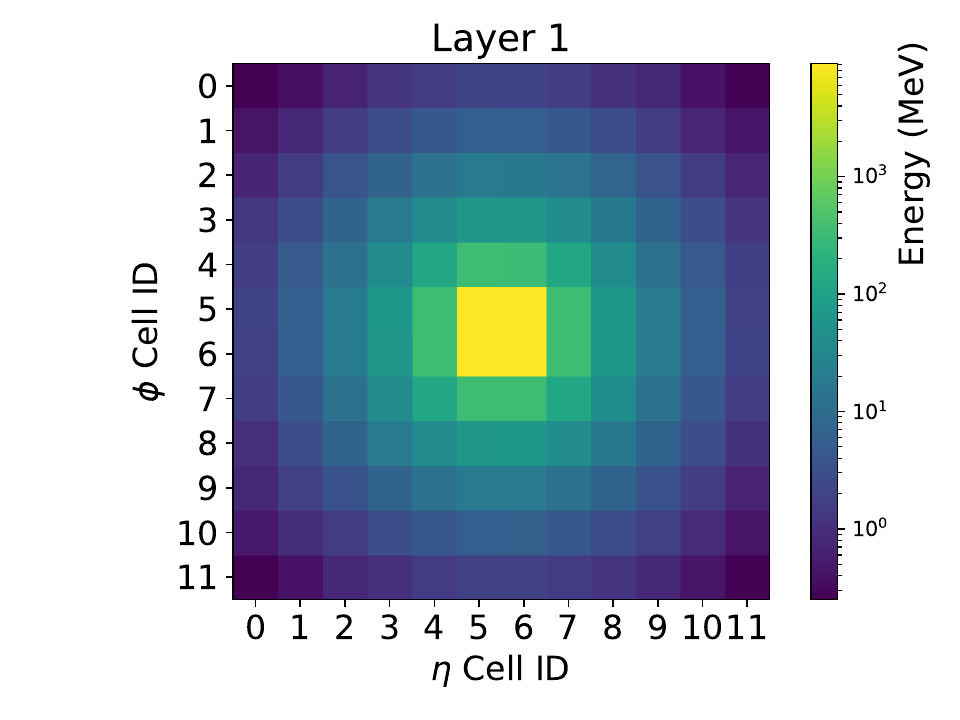}
    \includegraphics[width=0.3\textwidth]{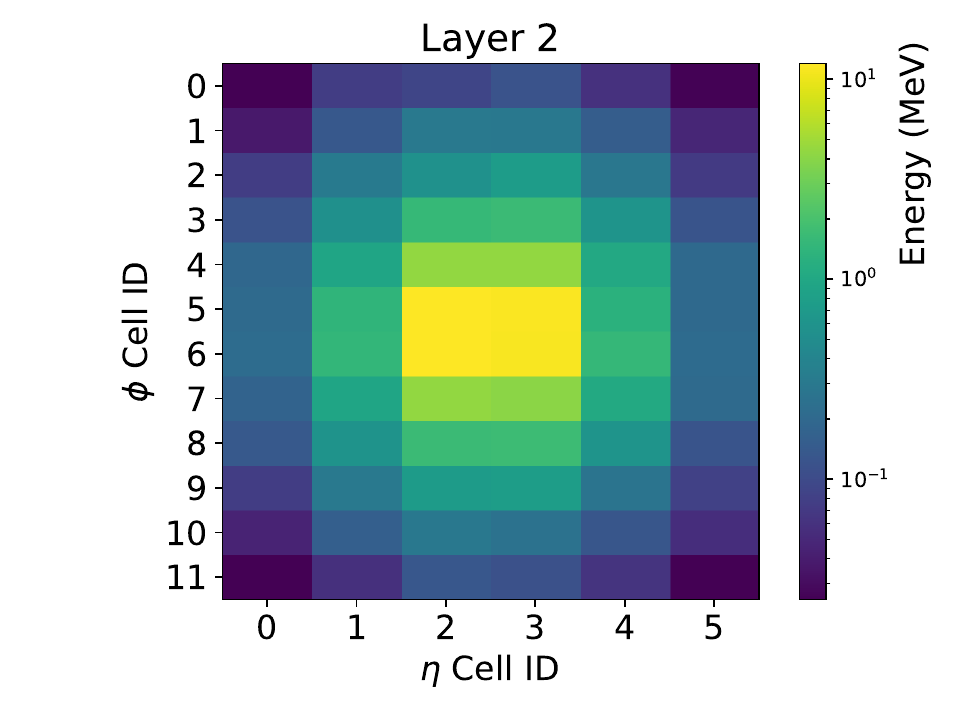}

    \includegraphics[width=0.3\textwidth]{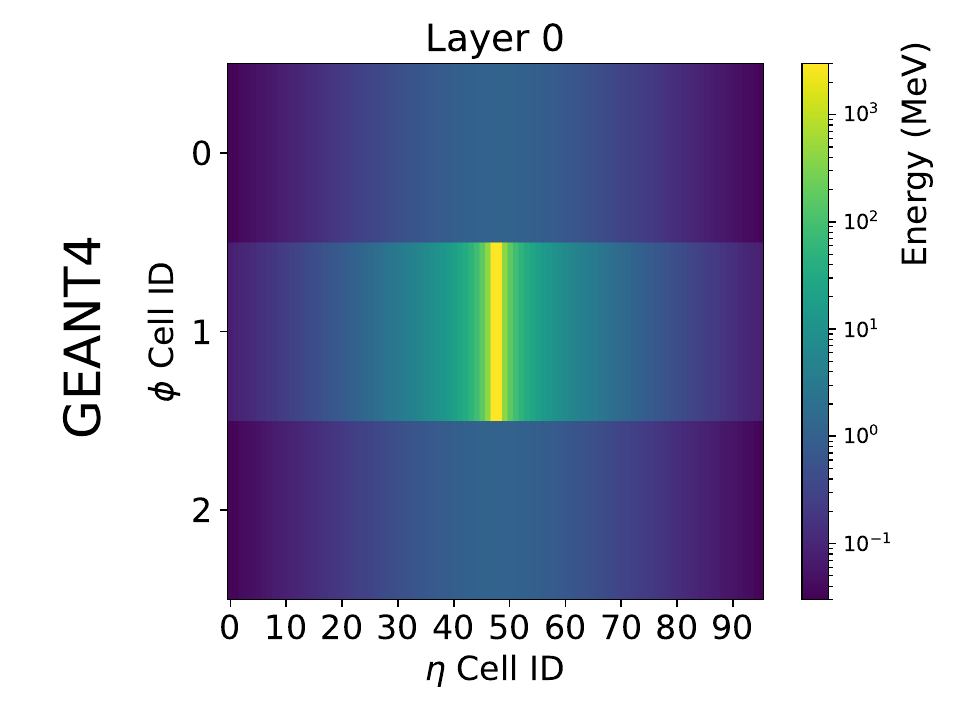}
    \includegraphics[width=0.3\textwidth]{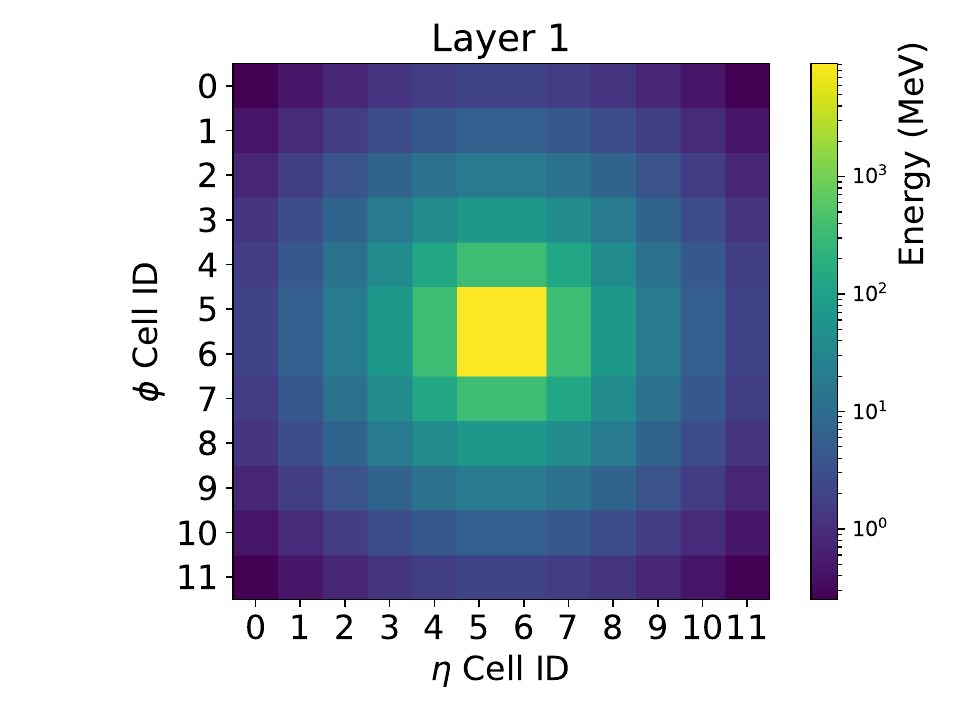}
    \includegraphics[width=0.3\textwidth]{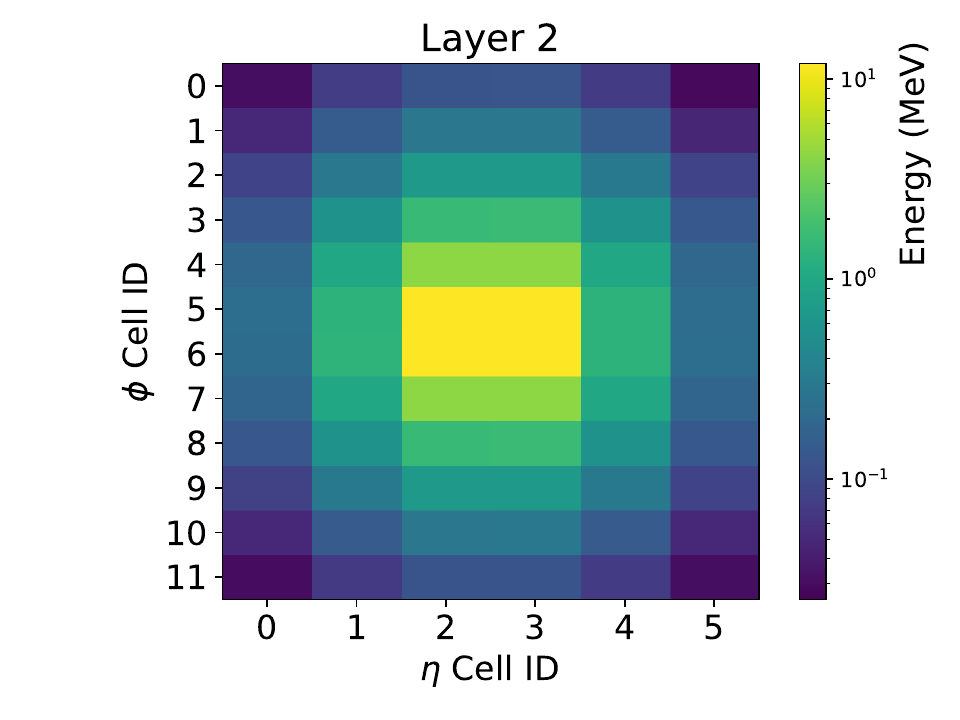}

    \includegraphics[width=0.3\textwidth]{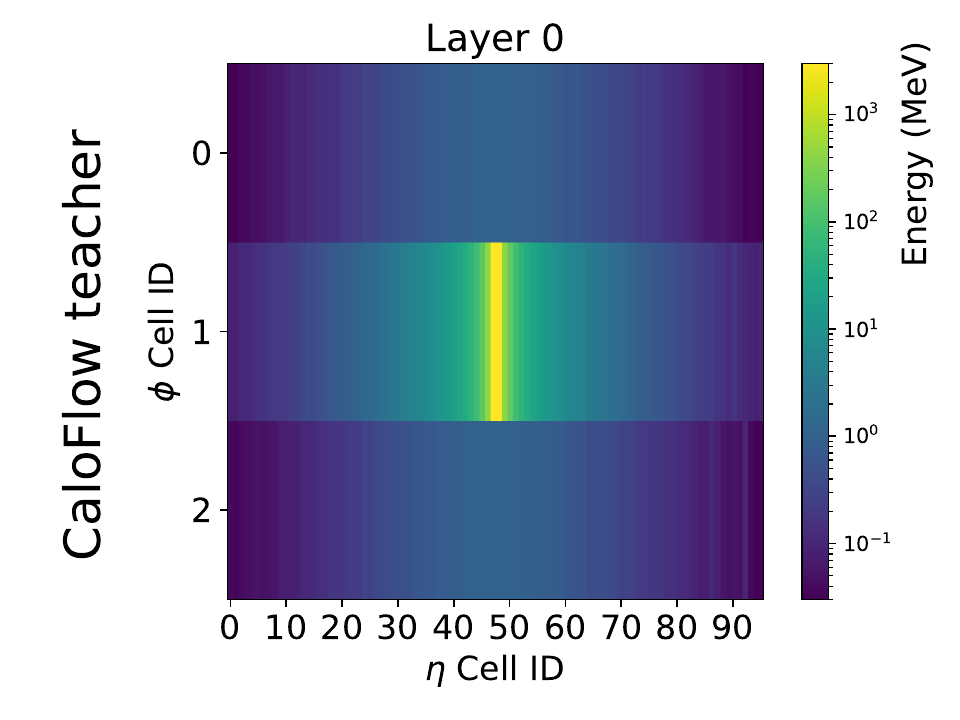}
    \includegraphics[width=0.3\textwidth]{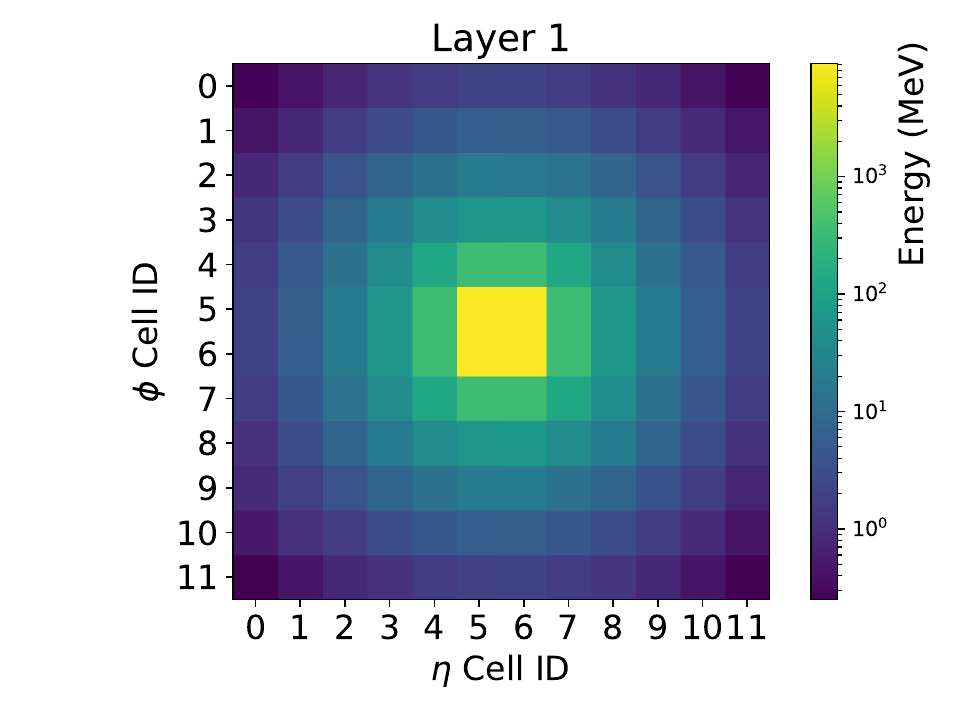}
    \includegraphics[width=0.3\textwidth]{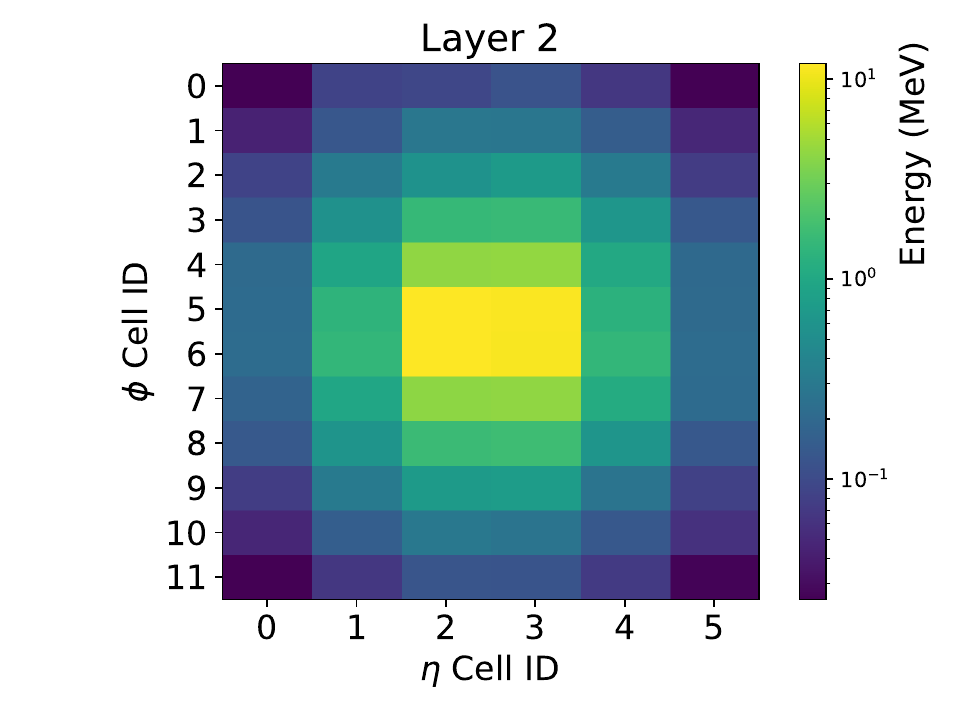}

    \caption{Average shower shapes for $e^{+}$. Columns are calorimeter layers 0 to 2, top row shows \cf\ student, center row \geant, and bottom row \cf\ teacher.}
    \label{fig:average.eplus}
\end{figure}

\begin{figure}[!ht]
    \centering

    \includegraphics[width=0.3\textwidth]{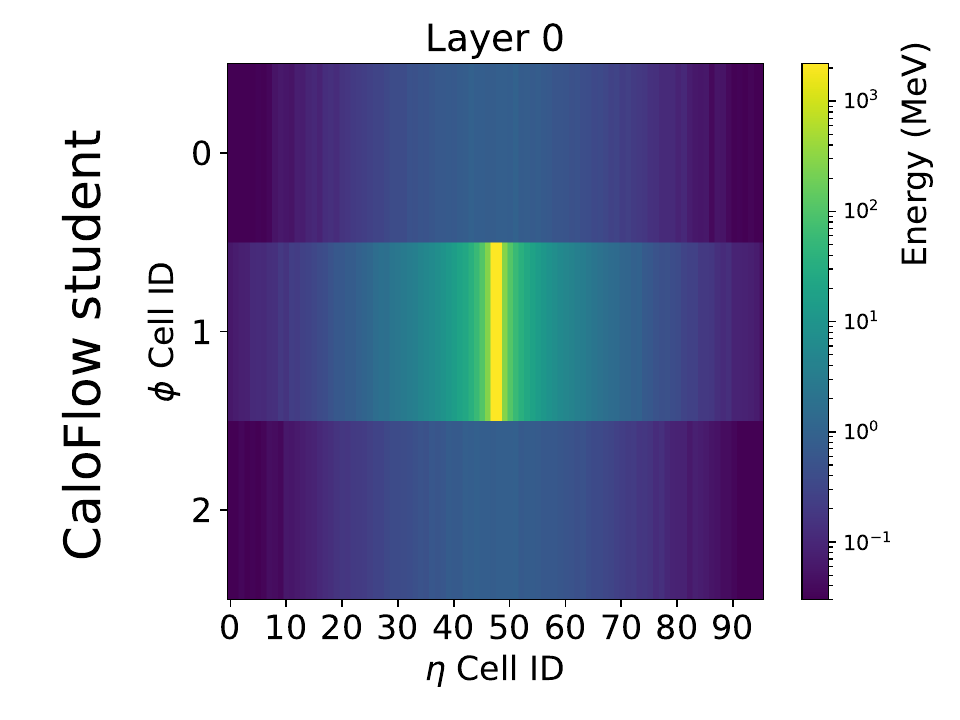}
    \includegraphics[width=0.3\textwidth]{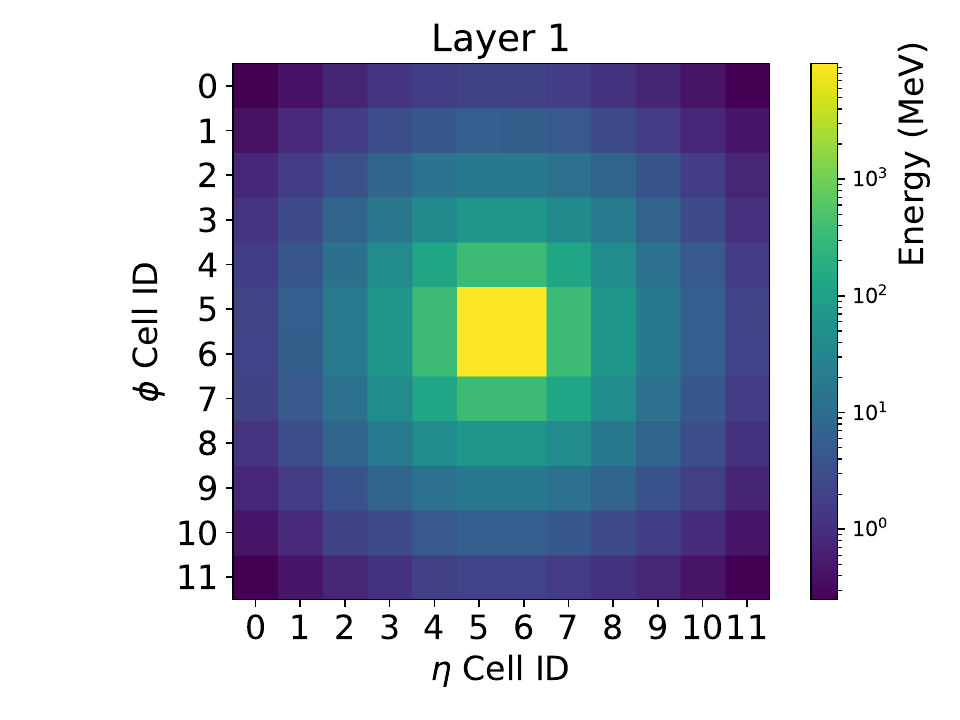}
    \includegraphics[width=0.3\textwidth]{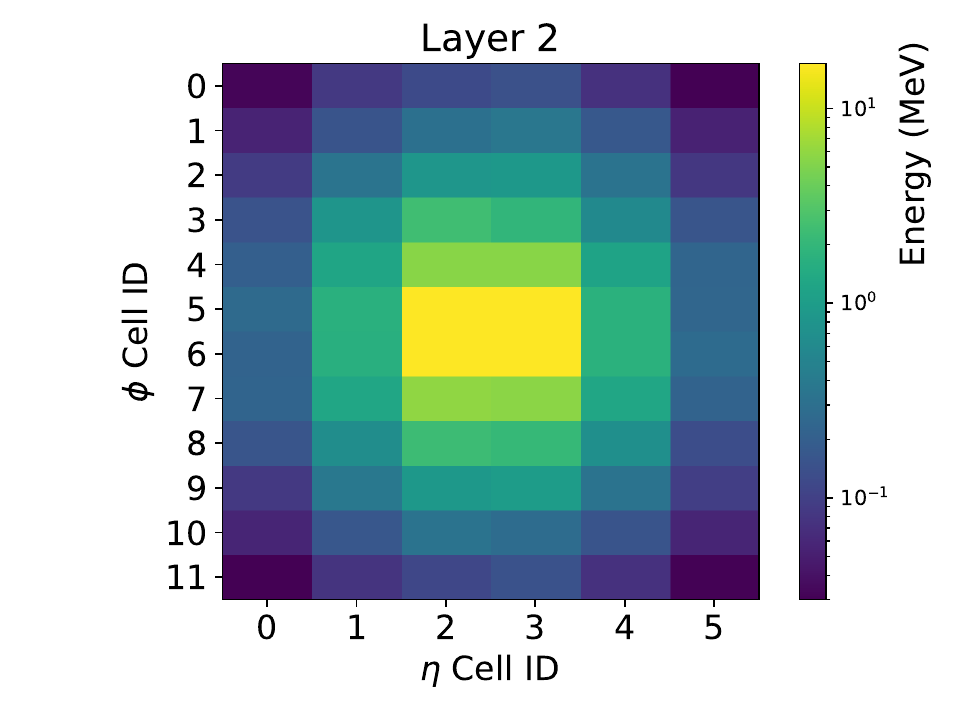}

    \includegraphics[width=0.3\textwidth]{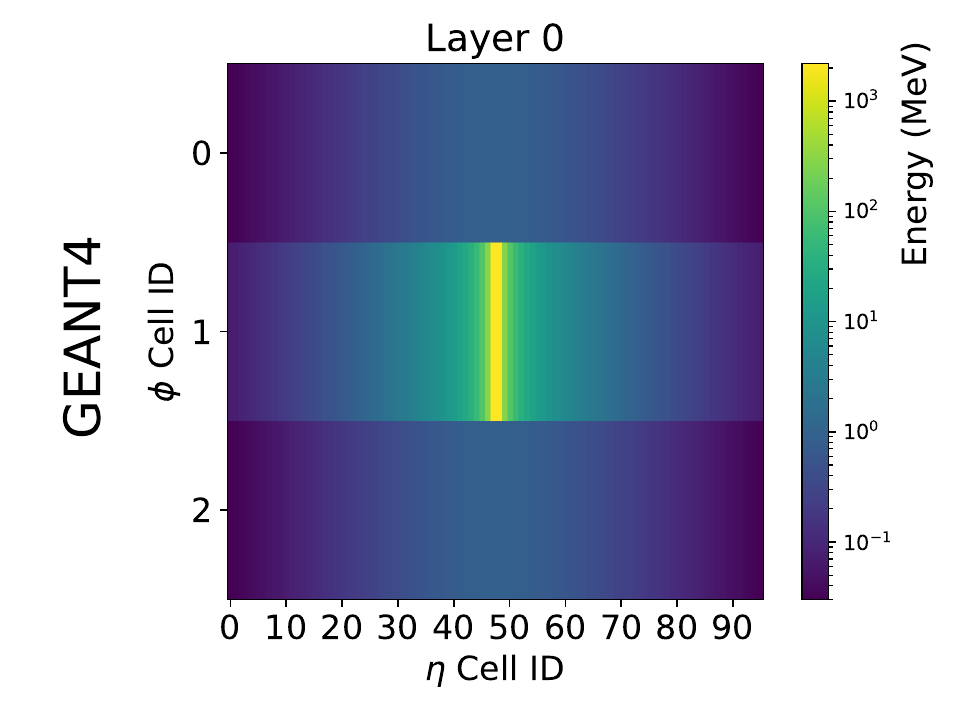}
    \includegraphics[width=0.3\textwidth]{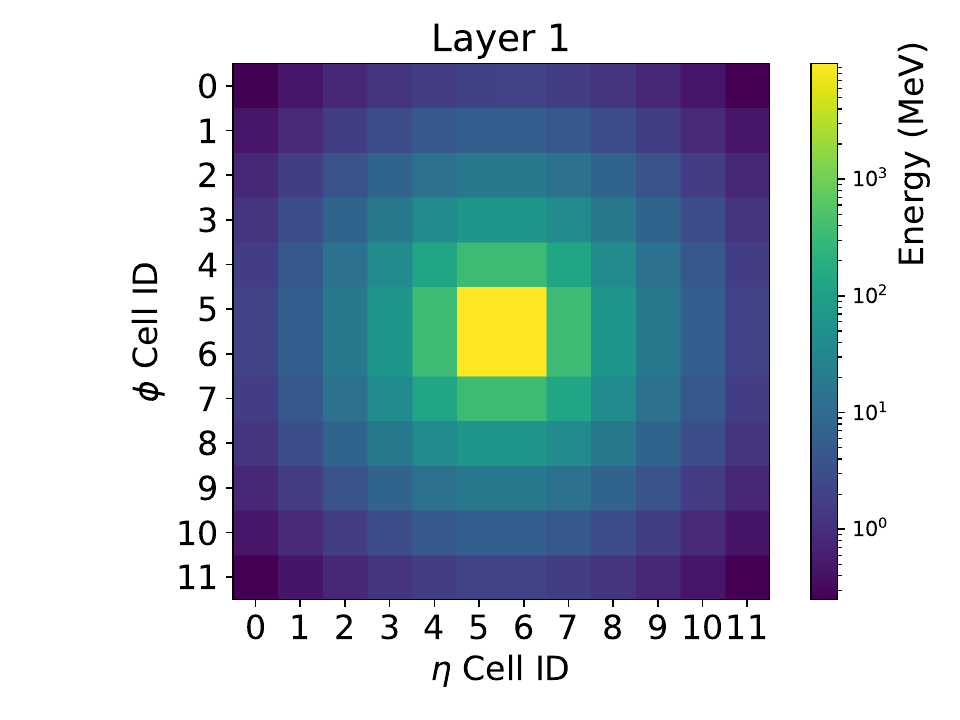}
    \includegraphics[width=0.3\textwidth]{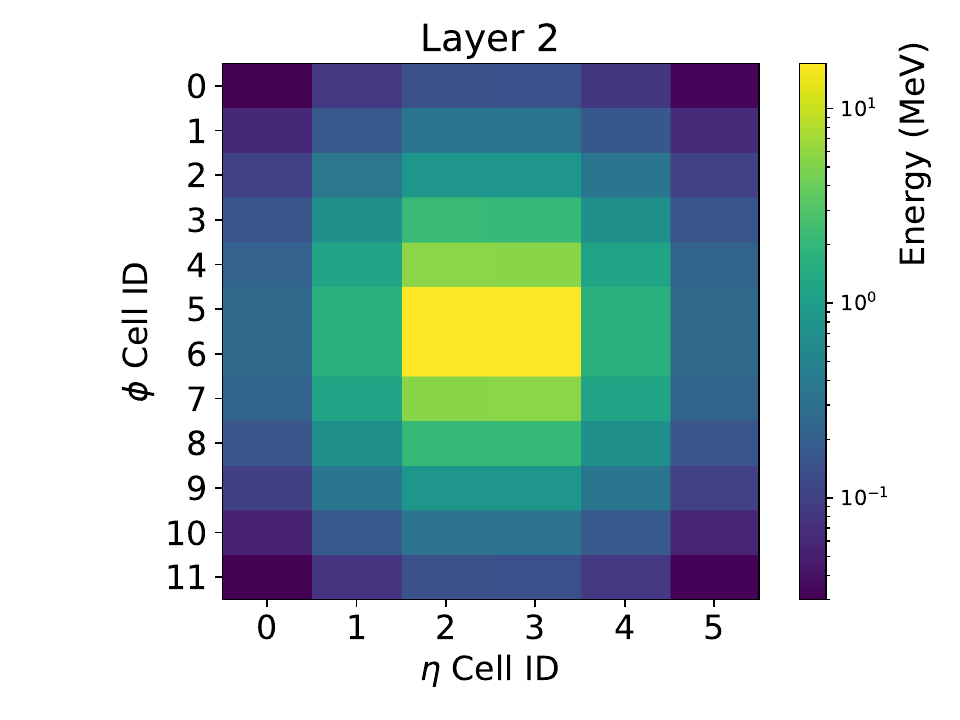}

    \includegraphics[width=0.3\textwidth]{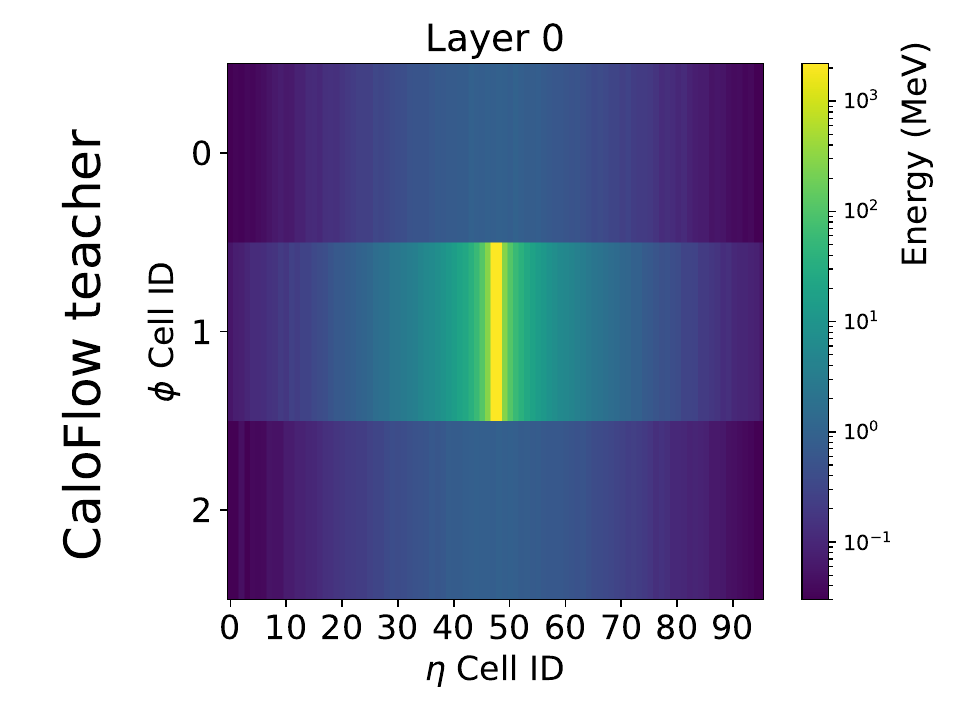}
    \includegraphics[width=0.3\textwidth]{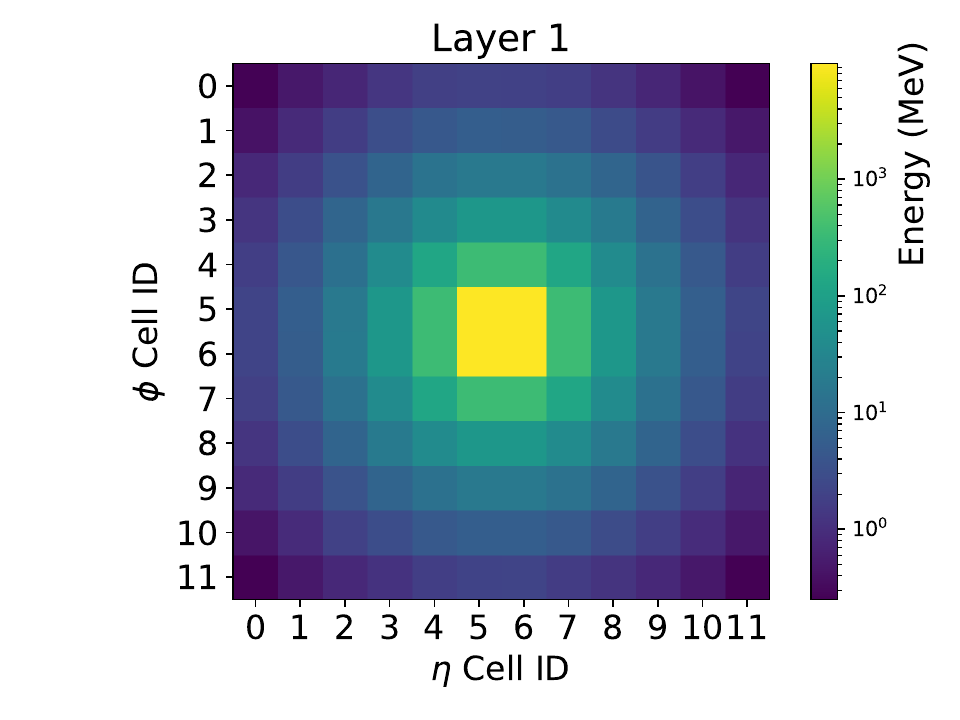}
    \includegraphics[width=0.3\textwidth]{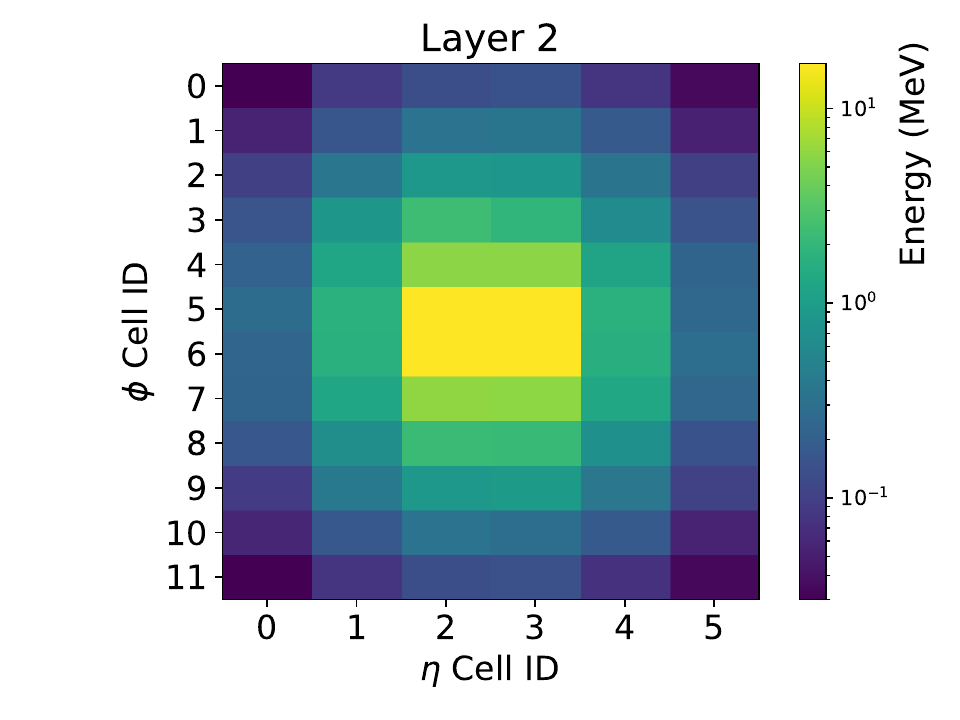}

    \caption{Average shower shapes for $\gamma$. Columns are calorimeter layers 0 to 2, top row shows \cf\ student, center row \geant, and bottom row \cf\ teacher.}
    \label{fig:average.gamma}
\end{figure}

\begin{figure}[!ht]
    \centering

    \includegraphics[width=0.3\textwidth]{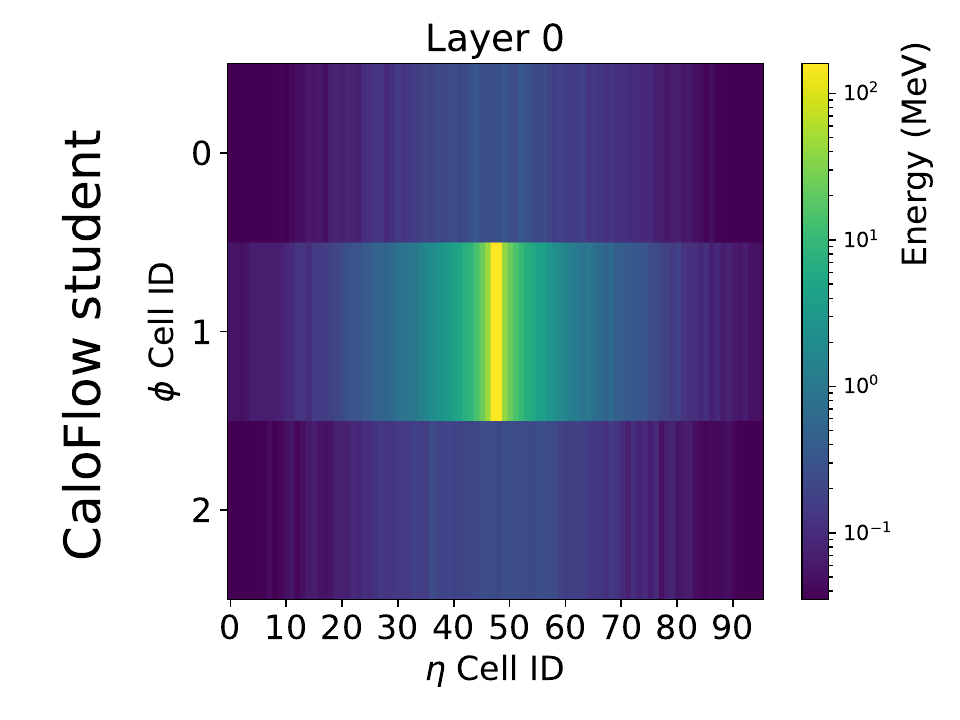}
    \includegraphics[width=0.3\textwidth]{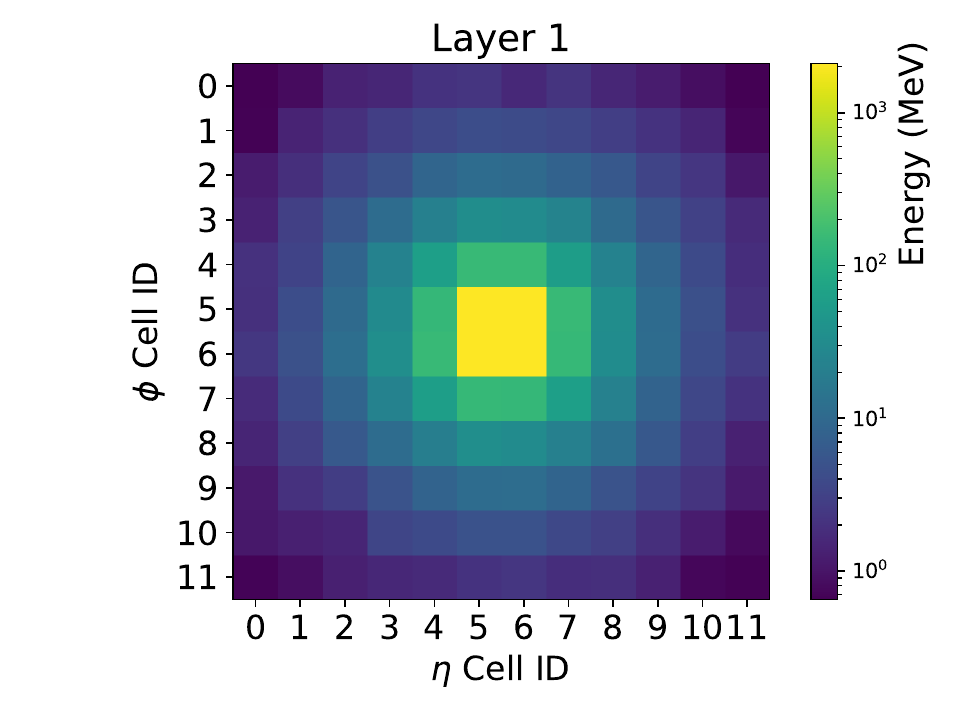}
    \includegraphics[width=0.3\textwidth]{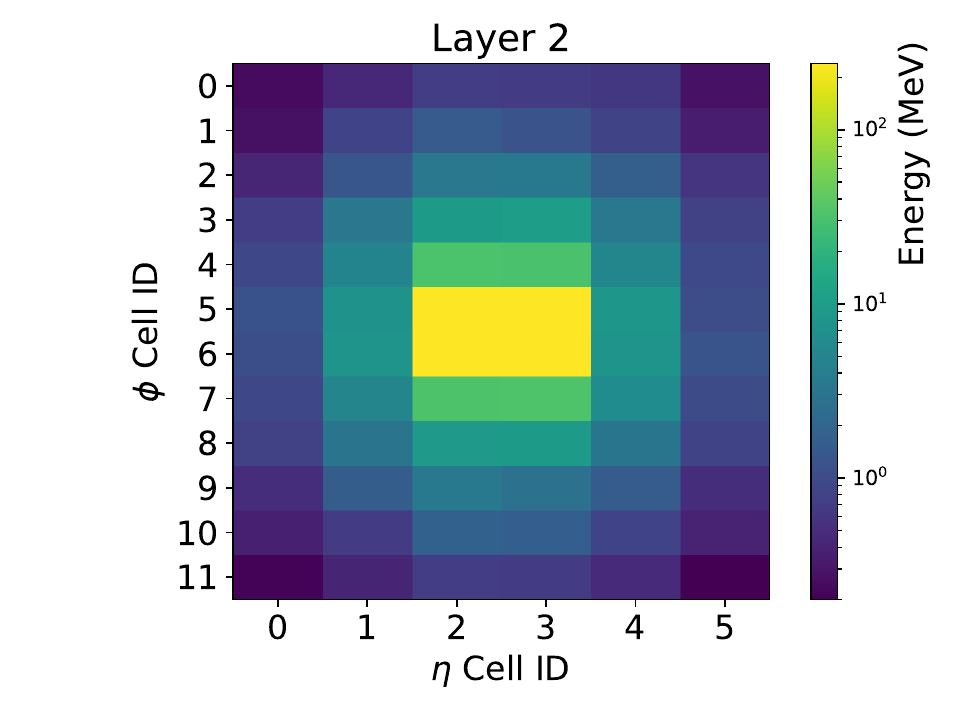}

    \includegraphics[width=0.3\textwidth]{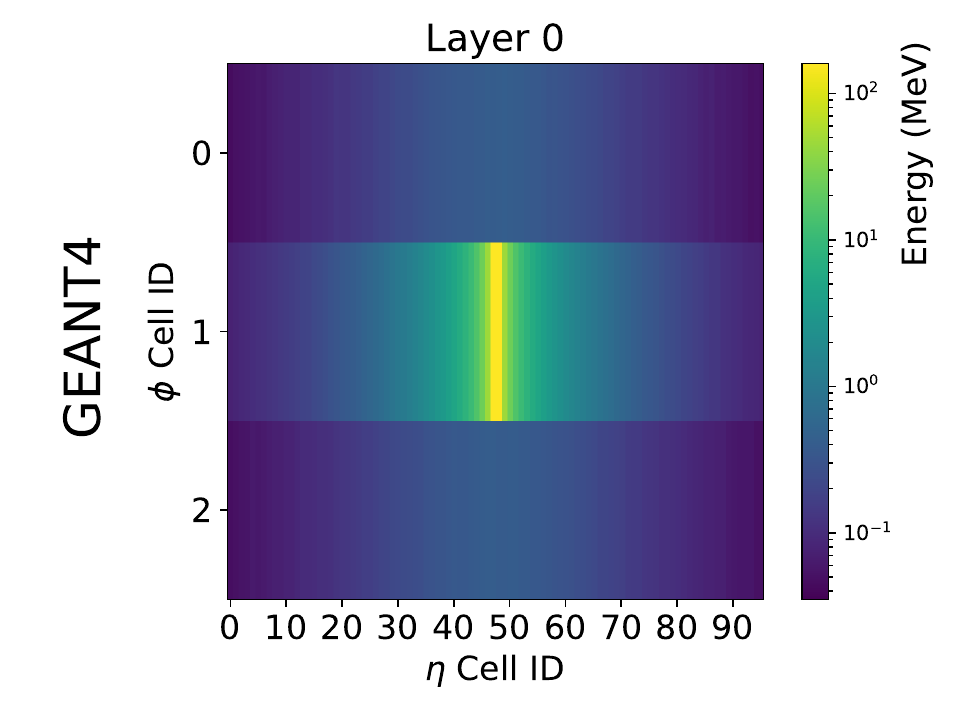}
    \includegraphics[width=0.3\textwidth]{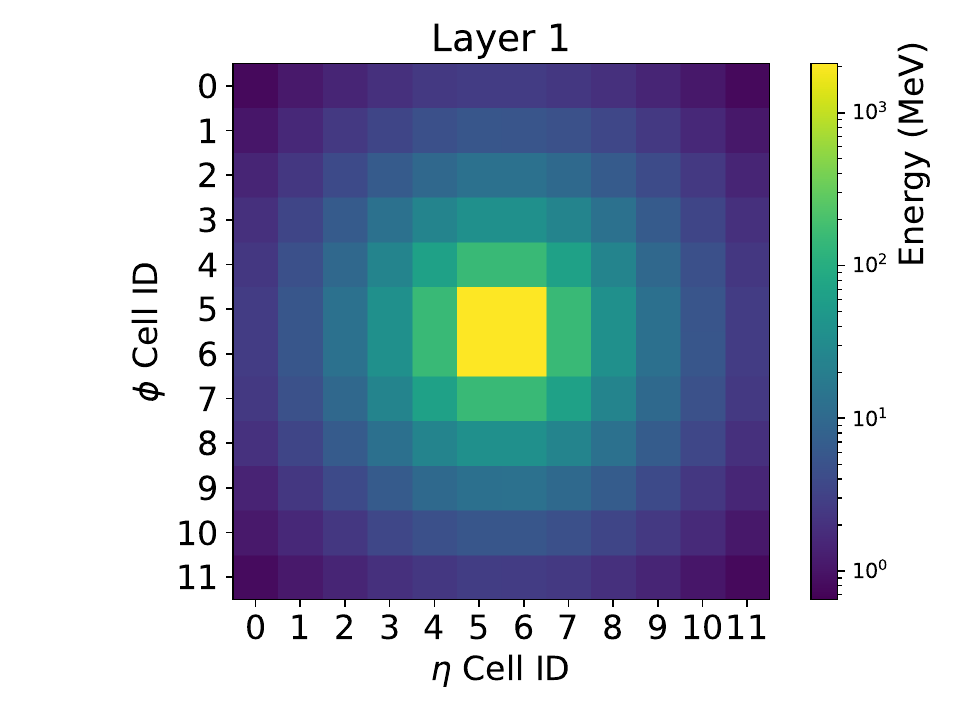}
    \includegraphics[width=0.3\textwidth]{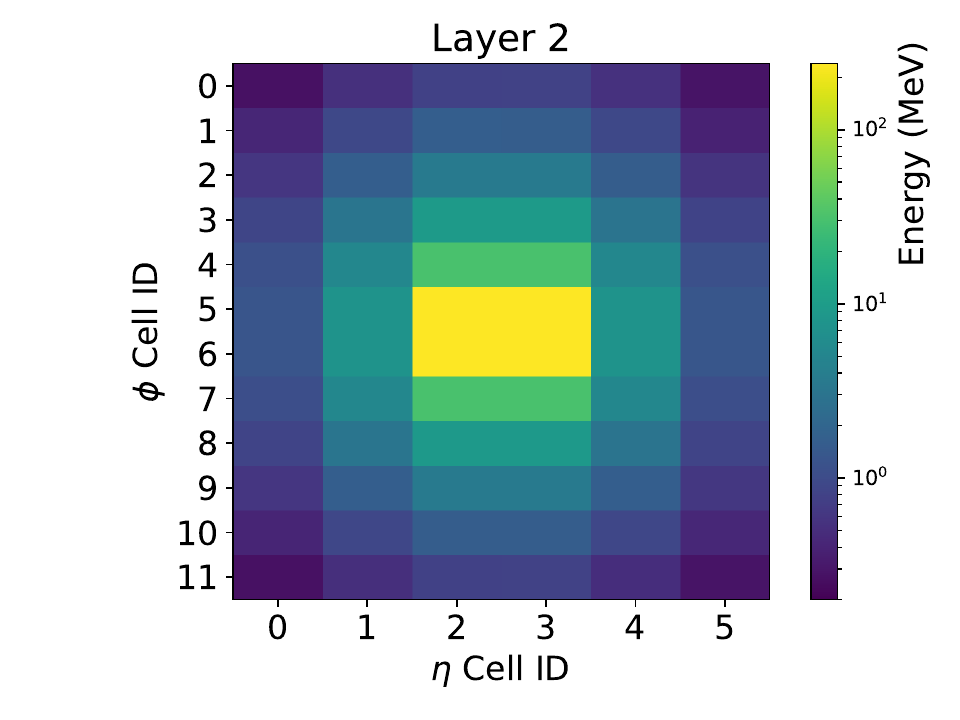}

    \includegraphics[width=0.3\textwidth]{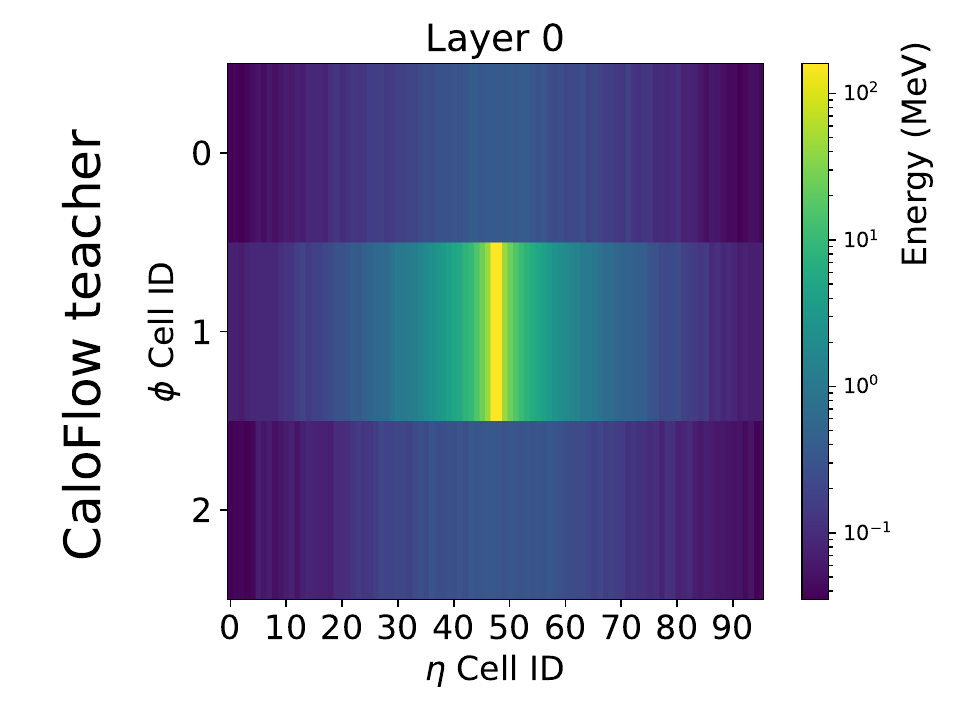}
    \includegraphics[width=0.3\textwidth]{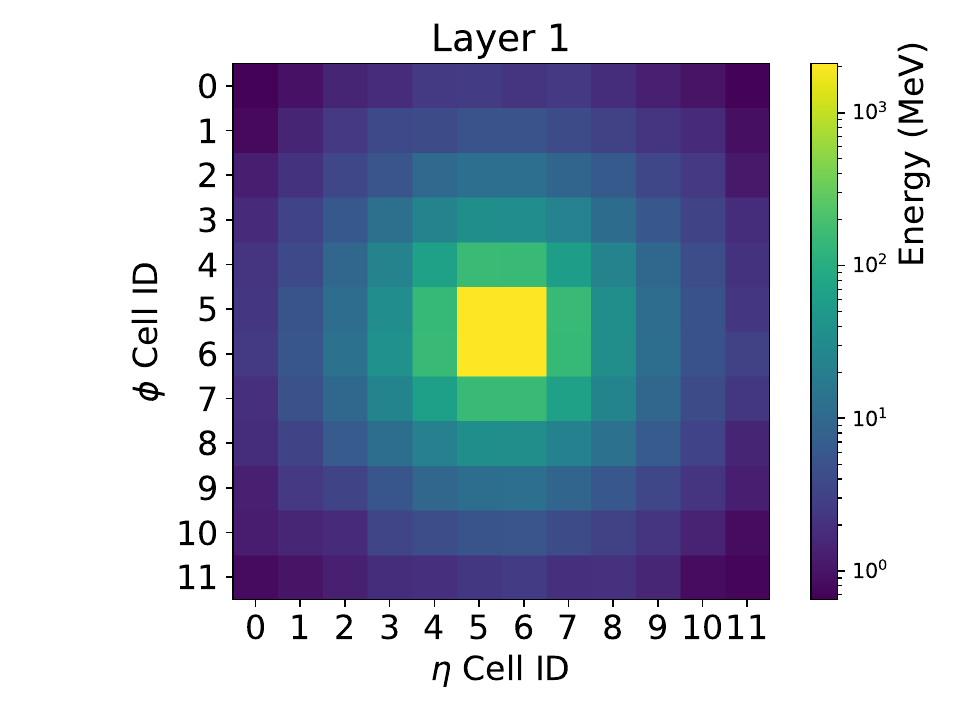}
    \includegraphics[width=0.3\textwidth]{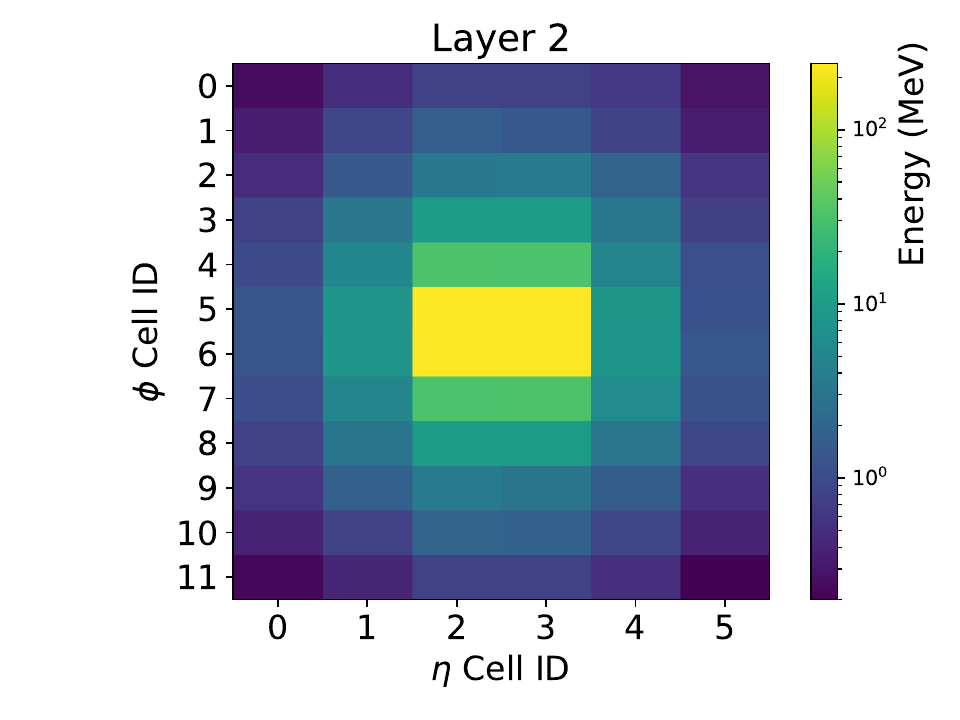}

    \caption{Average shower shapes for $\pi^{+}$. Columns are calorimeter layers 0 to 2, top row shows \cf\ student, center row \geant, and bottom row \cf\ teacher.}
    \label{fig:average.piplus}
  \end{figure}

Next we turn to the same qualitative comparisons of average and individual images that we conducted in \cite{Krause:2021ilc}, following \cite{Paganini:2017hrr,Paganini:2017dwg}. Shown in figs.~\ref{fig:average.eplus}, \ref{fig:average.gamma} and \ref{fig:average.piplus} are the average calorimeter shower images for $e^+$, $\gamma$ and $\pi^+$ respectively, for \geant, \cf\ teacher and \cf\ student. We see excellent agreement between all three; they are nearly indistinguishable by eye. There is no sign of mode-collapse.

  \clearpage

\subsection{Flow II histograms}
\label{sec:res2}

\begin{figure}[!ht]
    \centering
    \includegraphics[width=0.85\textwidth, trim=145 250 170 300, clip]{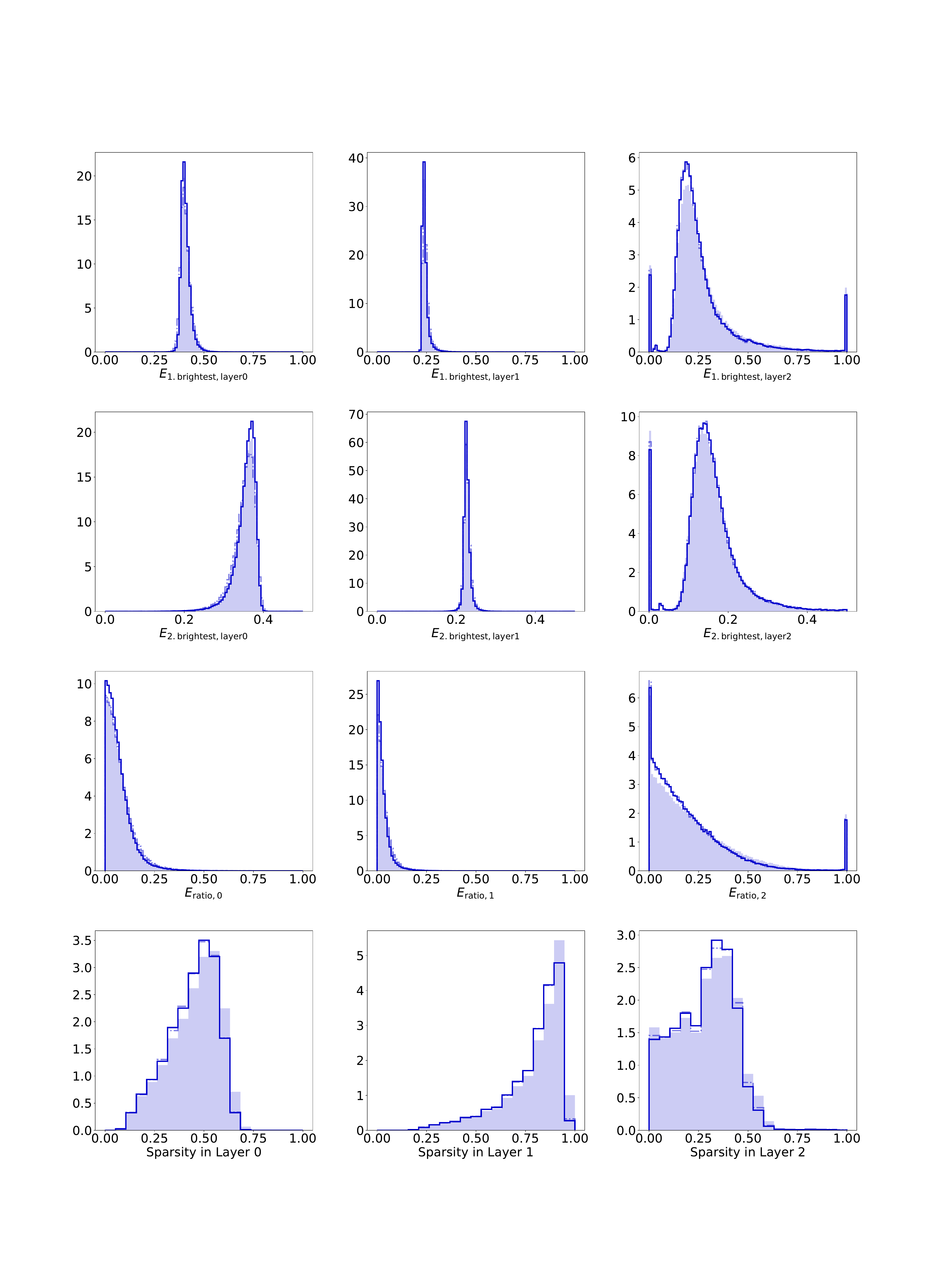}
    \includegraphics[width=0.75\textwidth]{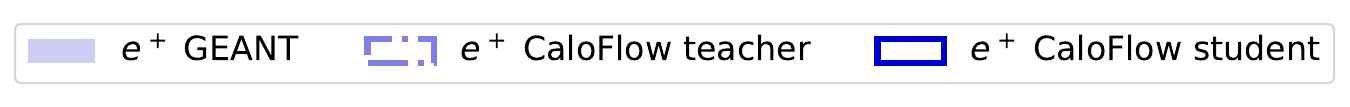}
    \caption{Distributions that are sensitive to Flow II for $e^{+}$. Top row: energy of brightest voxel compared to the layer energy; second row: energy of second brightest voxel compared to the layer energy; third row: difference of brightest and second brightest voxel, normalized to their sum; last row: sparsity of the showers. See~\cite{Krause:2021ilc} for detailed definitions.}
    \label{fig:flow2.voxel.histos.eplus}
\end{figure}

\begin{figure}[!ht]
    \centering
    \includegraphics[width=0.85\textwidth, trim=145 250 170 300, clip]{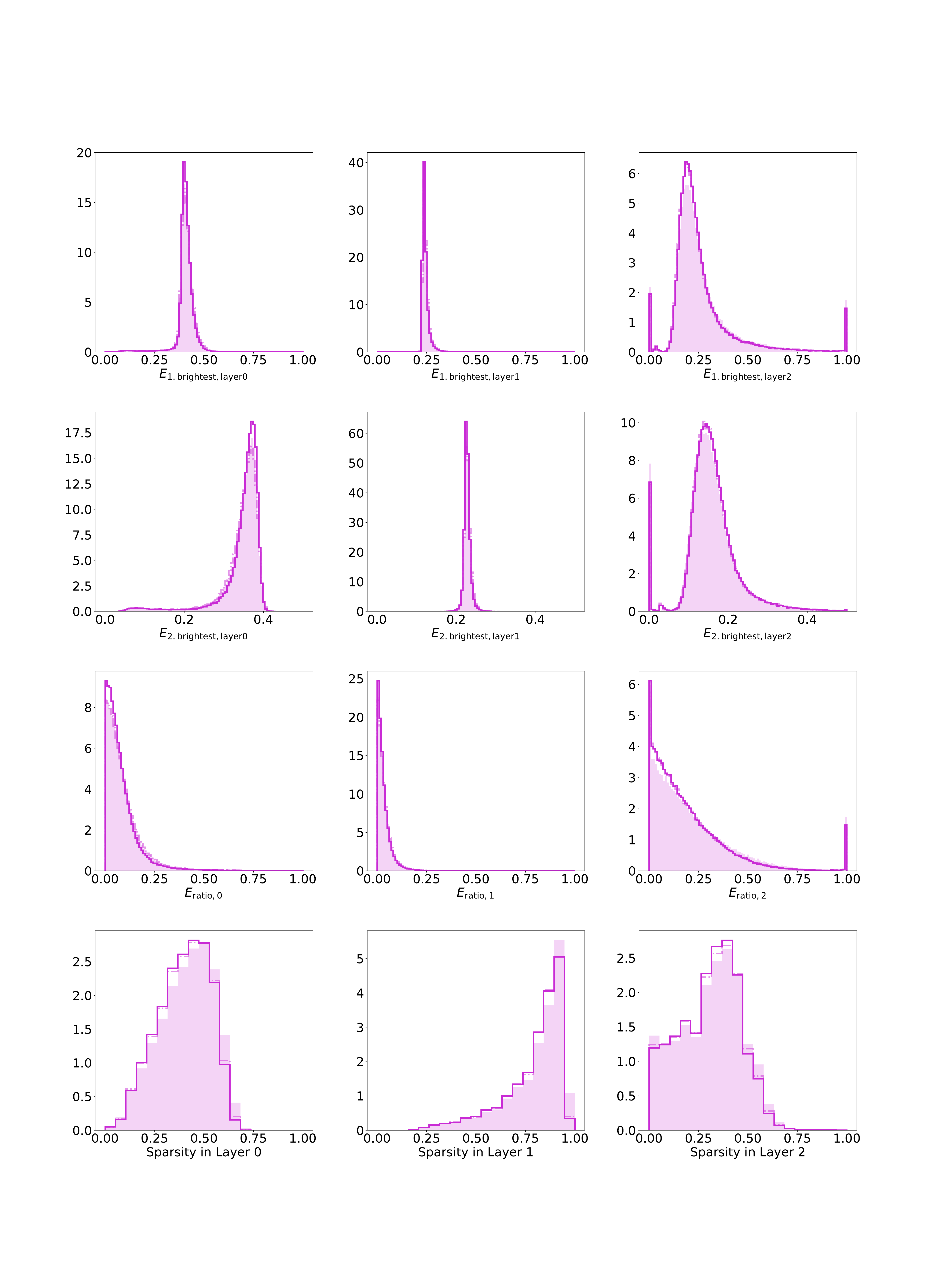}
    \includegraphics[width=0.75\textwidth]{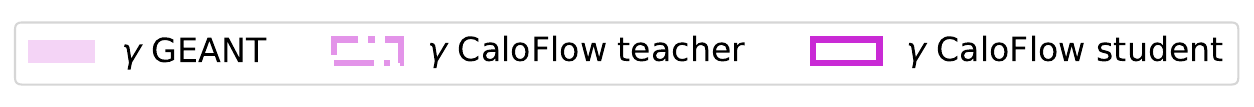}
    \caption{Distributions that are sensitive to Flow II for $\gamma^{+}$. Top row: energy of brightest voxel compared to the layer energy; second row: energy of second brightest voxel compared to the layer energy; third row: difference of brightest and second brightest voxel, normalized to their sum; last row: sparsity of the showers. See~\cite{Krause:2021ilc} for detailed definitions.}
    \label{fig:flow2.voxel.histos.gamma}
\end{figure}

\begin{figure}[!ht]
    \centering
    \includegraphics[width=0.85\textwidth, trim=145 250 170 300, clip]{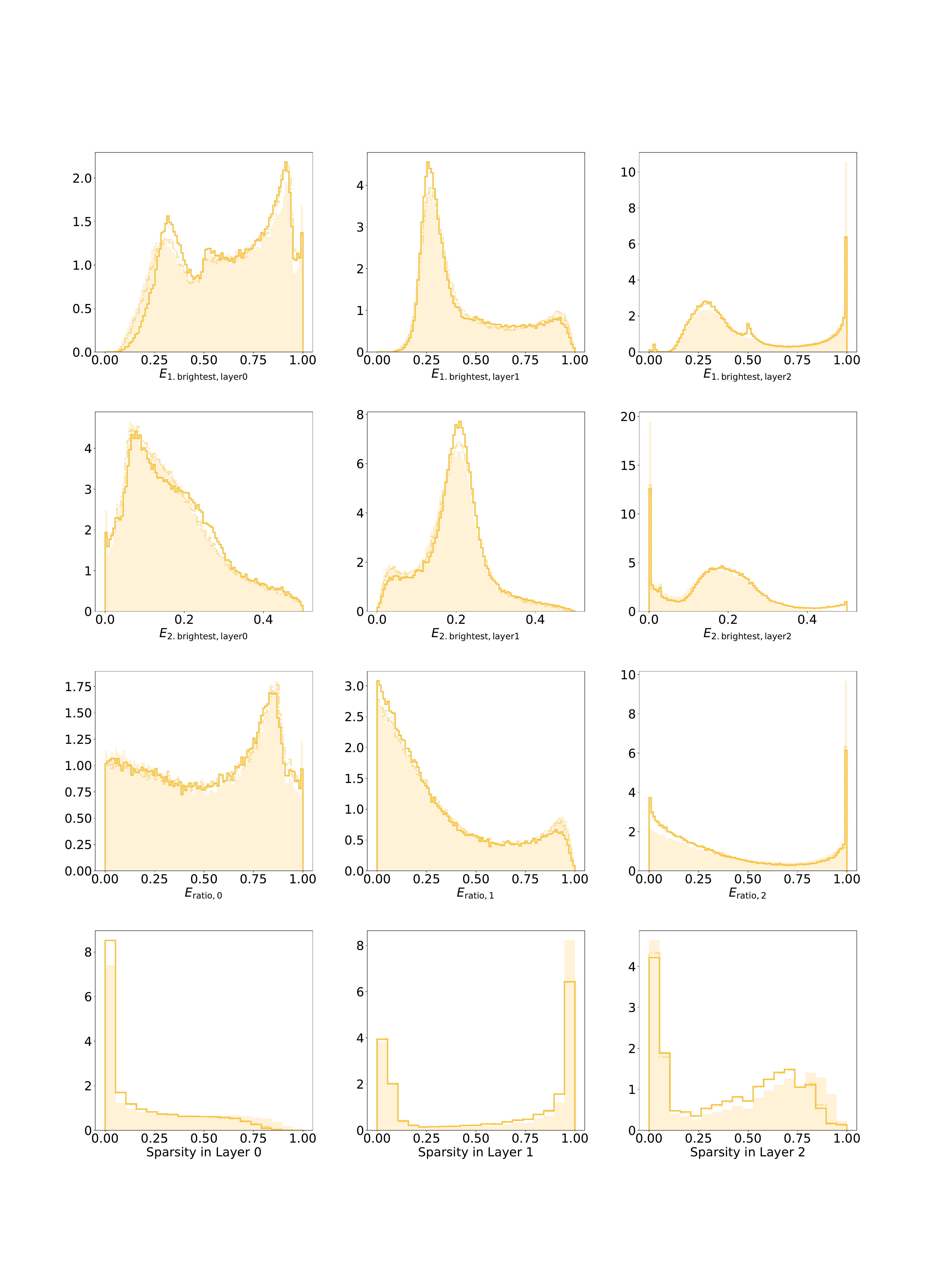}
    \includegraphics[width=0.75\textwidth]{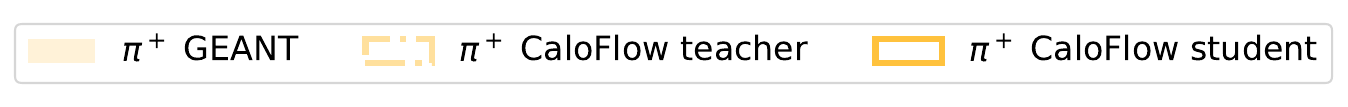}
    \caption{Distributions that are sensitive to Flow II for $\pi^{+}$. Top row: energy of brightest voxel compared to the layer energy; second row: energy of second brightest voxel compared to the layer energy; third row: difference of brightest and second brightest voxel, normalized to their sum; last row: sparsity of the showers. See~\cite{Krause:2021ilc} for detailed definitions.}
    \label{fig:flow2.voxel.histos.piplus}
  \end{figure}

\begin{figure}[!ht]
    \centering
    \includegraphics[width=0.85\textwidth, trim=145 150 170 200, clip]{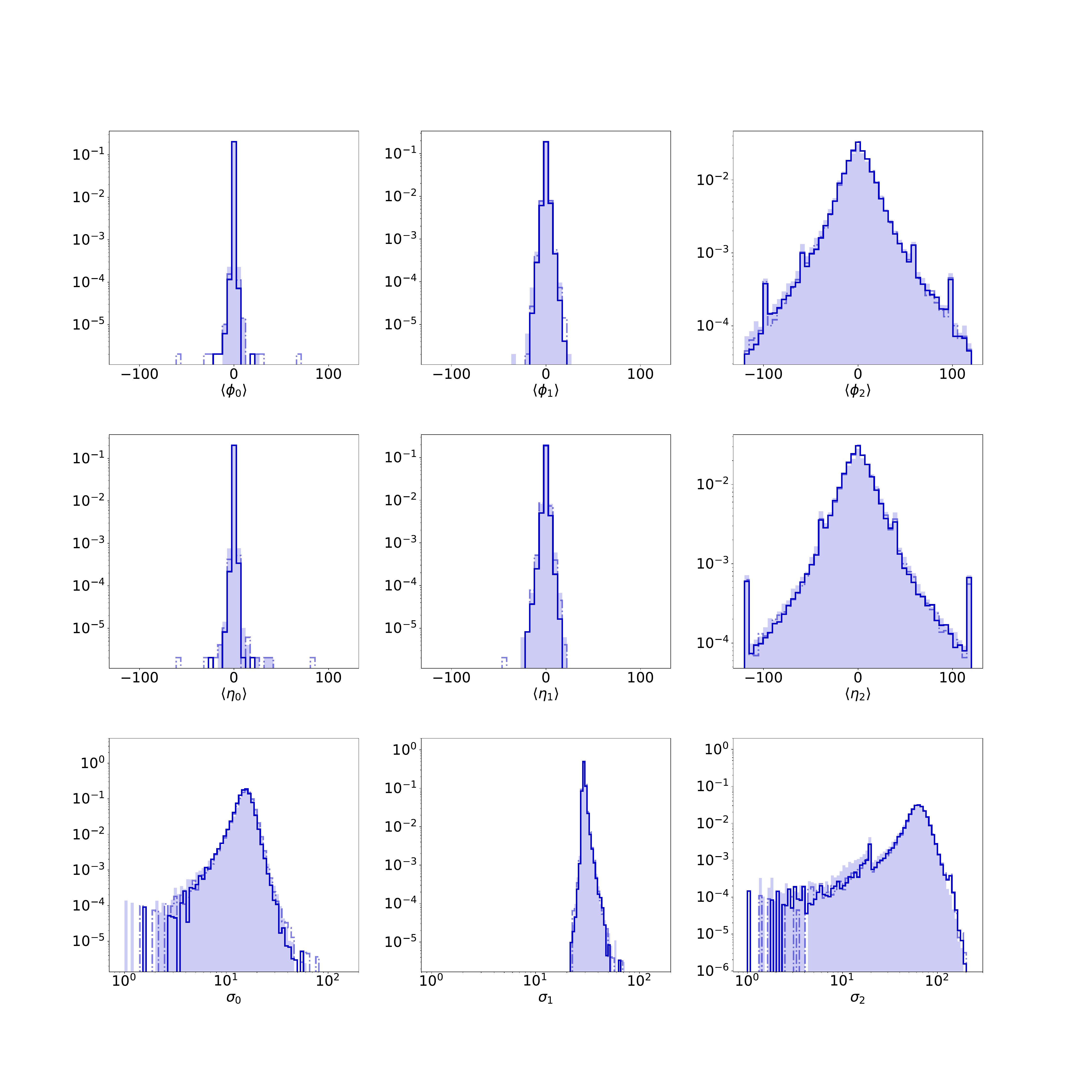}
    \includegraphics[width=0.75\textwidth]{figures/eplus/legend.pdf}
    \caption{Further distributions that are sensitive to Flow II for $e^{+}$, as learned by Flow II. Top and center row show the location of the deposition centroid in $\phi$ and $\eta$ direction; the bottom row shows the standard deviation of the $\eta$ centroid.}
    \label{fig:flow2.shower.histos.eplus}
\end{figure}

\begin{figure}[!ht]
    \centering
    \includegraphics[width=0.85\textwidth, trim=145 150 170 200, clip]{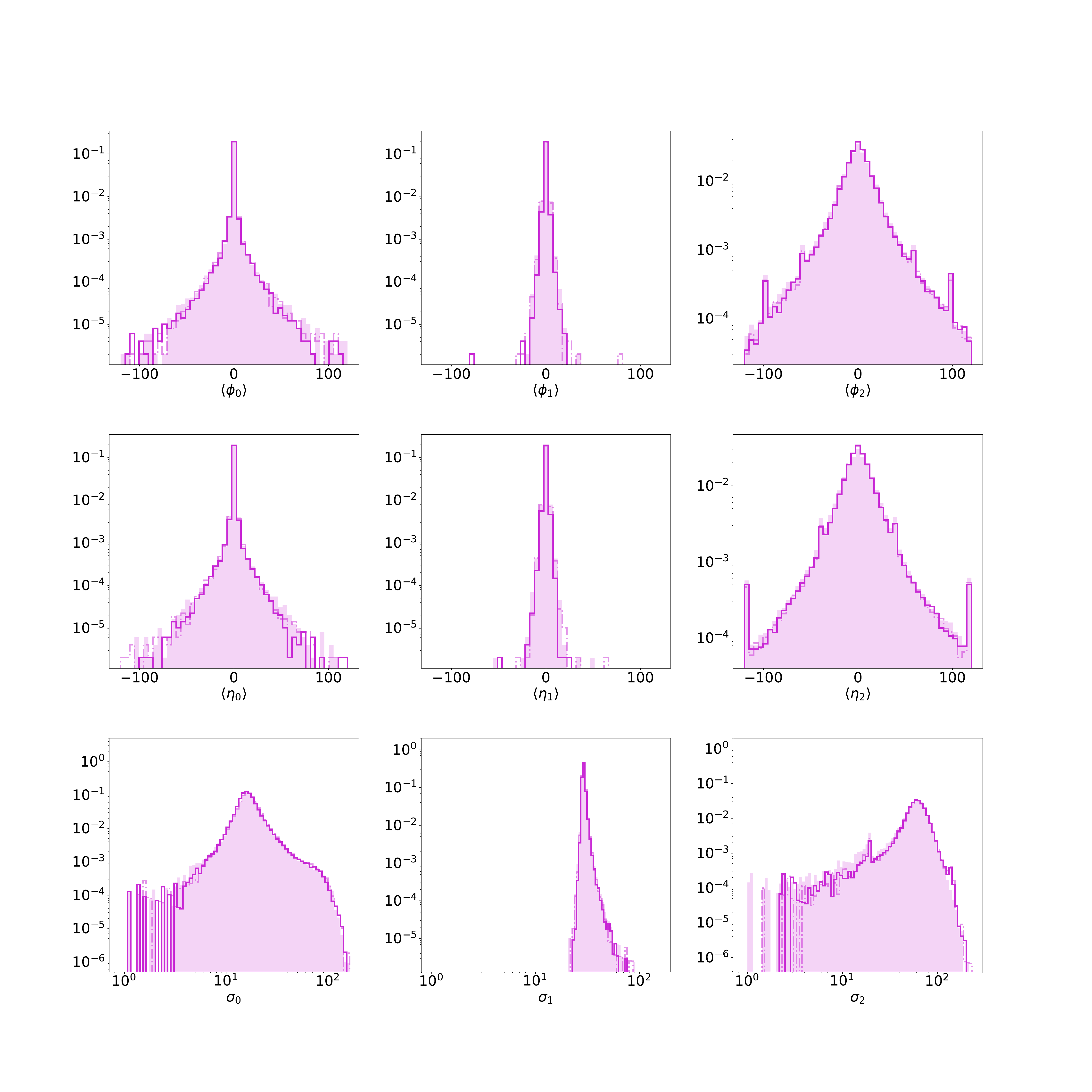}
    \includegraphics[width=0.75\textwidth]{figures/gamma/legend.pdf}
    \caption{Further distributions that are sensitive to Flow II for $\gamma$, as learned by Flow II. Top and center row show the location of the deposition centroid in $\phi$ and $\eta$ direction; the bottom row shows the standard deviation of the $\eta$ centroid.}  
    \label{fig:flow2.shower.histos.gamma}
\end{figure}

\begin{figure}[!ht]
    \centering
    \includegraphics[width=0.85\textwidth, trim=145 150 170 200, clip]{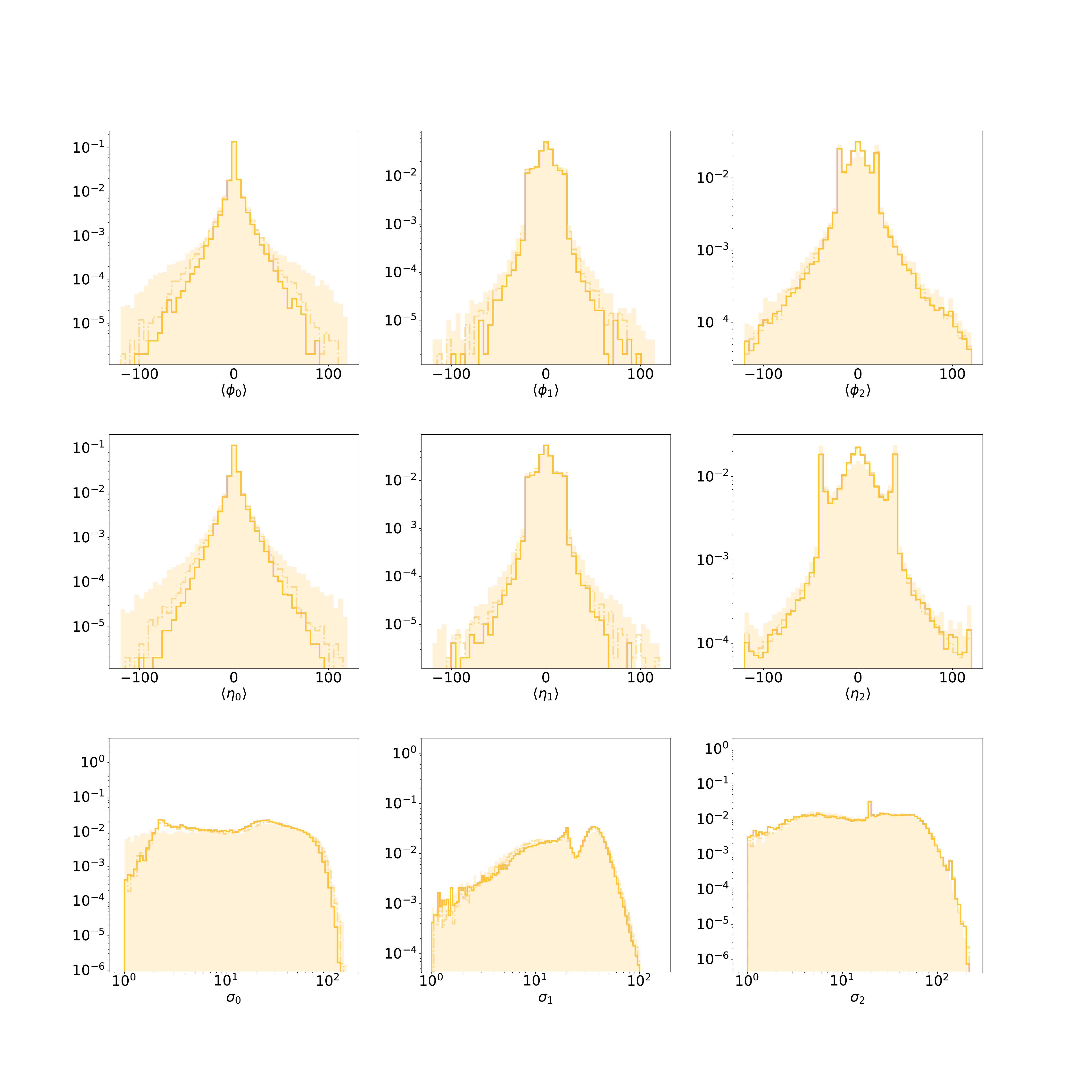}
    \includegraphics[width=0.75\textwidth]{figures/piplus/legend.pdf}
    \caption{Further distributions that are sensitive to Flow II for $\pi^{+}$, as learned by Flow II. Top and center row show the location of the deposition centroid in $\phi$ and $\eta$ direction; the bottom row shows the standard deviation of the $\eta$ centroid.}
    \label{fig:flow2.shower.histos.piplus}
\end{figure}

Figures~\ref{fig:flow2.voxel.histos.eplus}--\ref{fig:flow2.shower.histos.piplus} show histograms of the same features as in~\cite{Krause:2021ilc} relevant for Flow II, for the three different particle types.\footnote{We do not show histograms that are only sensitive to Flow I, as Flow I of \cfvI\ and \cfvII\ are identical and the only differences in histograms would be of statistical nature.} These are the two brightest voxels in each layer, the difference of those two divided by their sum (called $E_{\text{ratio}}$), the fraction of voxels with an energy deposition (called sparsity), the centroid in $\phi$ and $\eta$ direction, and the standard deviation of the $\eta$ centroid (called $\sigma_{i}$); see~\cite{Krause:2021ilc} for more details. Each histogram compares the \geant\ reference sample with the Flow II teacher (taken from~\cite{Krause:2021ilc}) and the new Flow II student. We again see in nearly all cases that the teacher and student are basically indistinguishable from one another. The largest differences between student and teacher are visible in the distributions of the brightest and second brightest pixels of layer 0 and layer 1 for $\pi^+$. Smaller differences between student and teacher can be seen in $E_{2,{\rm brightest},{\rm layer0}}$ and $E_{{\rm ratio},0}$ for $e^+$ and $\gamma$. Curiously, in one histogram ($E_{2,{\rm brightest},{\rm layer0}}$ for $e^+$) the student actually matches the \geant\ reference {\it better} than the teacher. This is possible if the teacher is off from the data, and the student hasn't fully converged to the teacher, leading to an accidentally better agreement with the data. 
Finally, we observe that for $\pi^{+}$-showers, there are also some small differences in the energy-weighted shower mean of the student in the top two rows of fig.~\ref{fig:flow2.shower.histos.piplus}; these appear to be slighly narrower than their teacher's equivalent. However, these differences are off-peak and subleading.

In addition to these histograms, app.~\ref{app:nearest} collects nearest neighbor comparisons of samples from \cfvII\ and \geant. As already in~\cite{Krause:2021ilc}, we do not observe any sign of mode collapse in these.

\clearpage

\subsection{Classifier metrics}
\label{sec:cls}
Next, we exhibit the result of the classifier metric introduced in~\cite{Krause:2021ilc}. It gauges the quality of the generative model through the score of a binary classifier trained to discriminate between the reference data sample and the generative model sample, approximating the Neyman-Pearson classifier. Unlike in \cite{Krause:2021ilc}, here we focus only on a simple DNN classifier trained on either all of the pixels of the calorimeter shower or on a set of high-level features.\footnote{The list of high-level features, DNN architecture, and training procedure is the same as in~\cite{Krause:2021ilc}, we train for 150 epochs with a learning rate of $10^{-3}$.} For simplicity we do not consider a CNN classifier (which takes considerably longer to train and was a less sensitive metric than the DNN in \cite{Krause:2021ilc}), but we do consider the same two preprocessing approaches for the low-level features. These are: ``unnormalized'', i.e.\ using the showers as they were generated as input to the classifier; and ``normalized'', i.e.\ using showers that are normalized such that they sum to 1 in each calorimeter layer as input to the classifier. In addition to the energy depositions of each voxel, we give the incident energy and the energy deposition per calorimeter layer to the classifier. The detailed list of high-level features and their preprocessing can be found in~\cite{Krause:2021ilc}. Before the final evaluation, we calibrate the classifiers using isotonic regression~\cite{2017arXiv170604599G} of {\tt sklearn}~\cite{scikit-learn} based on the validation dataset, see~\cite{Krause:2021ilc} for more details.

\renewcommand{\arraystretch}{1.5}
\begin{table*}[!t]
\caption{AUC and JSD metrics for the classification of \geant\ vs.\ \cf\ student showers (lower numbers are better). Classifiers were trained on each particle type ($e^{+}$, $\gamma$, $\pi^{+}$) separately. All entries show mean and standard deviation of 10 classifier re-trainings on the same sample and are rounded to 3 digits. For comparison, we also give the classifier scores of the \cf\ teacher of~\cite{Krause:2021ilc}.} 
\label{tab:classifier}
\begin{center}
\begin{tabular}{|c|c|c|c|}
  \hline
  \multicolumn{2}{|c|}{\multirow{2}{*}{AUC / JSD}} & \geant\ vs.& \geant\ vs.\\
  \multicolumn{2}{|c|}{} & \cfvII\ (student) & \cfvI\ (teacher)~\cite{Krause:2021ilc} \\
  \hline
  \multirow{3}{*}{$e^{+}$}&unnormalized & 0.786(7) / 0.201(11)&  0.859(10) / 0.365(14)  \\
  \cline{2-4}
  &  normalized & 0.824(4) / 0.257(8)  & 0.870(2) / 0.378(5) \\
  \cline{2-4}
  & high-level & 0.762(3) / 0.164(5) & 0.795(1) / 0.229(3) \\
  \hline
  \multirow{3}{*}{$\gamma$}&unnormalized & 0.758(14) / 0.162(18)  & 0.756(48) / 0.174(68) \\
  \cline{2-4}
  &  normalized & 0.760(3) / 0.158(4) & 0.796(2) / 0.216(4) \\ 
  \cline{2-4}
  & high-level & 0.739(2) / 0.139(3) & 0.727(2) / 0.131(3) \\
  \hline
  \multirow{3}{*}{$\pi^{+}$}&unnormalized & 0.729(2) / 0.144(3)	&  0.649(3) / 0.060(2)  \\
  \cline{2-4}
  &  normalized & 0.807(1) / 0.230(3) & 0.755(3) / 0.153(3) \\
  \cline{2-4}
  & high-level & 0.893(2) / 0.410(5) & 0.888(1) / 0.401(4) \\
  \hline
\end{tabular}
\end{center}
\end{table*}
\renewcommand{\arraystretch}{1}

We see in tab.~\ref{tab:classifier} that in all cases, the classifier scores of the student are in line with those of the teacher, sometimes slightly worse, and sometimes even slightly better. Most importantly, they are always significantly different from unity which indicates that they always remain much higher-fidelity than the GAN. (Recall, in~\cite{Krause:2021ilc}, we showed the DNN trained on GAN vs.\ \geant\ achieved AUC$=1$ for all three particle types.) The fact that the student quality surpasses the teacher's in some cases can be explained by the observation we made in section~\ref{sec:res2}: Some features that are not perfectly modeled by the teacher can get accidentally better in a student that does not exactly follow the teacher.

\subsection{Timing benchmarks}
\label{sec:timing}

Finally, we come to the main {\it raison d'\^etre} for the student IAF: realizing the factor of $d\sim 500$ gain in sampling speed compared to the MAF. 
We summarize training and generation times of \cfvI, \cfvII, \cg, and \geant\ in table~\ref{tab:timing}. Timings are evaluated on our \textsc{Titan V} GPU, except for the \geant\ runtime, which is taken from~\cite{Paganini:2017dwg}. The training of the student is understood as being in addition to training the teacher. The difference in generation times for different batch sizes in \cg\ is due to {\tt Keras-TensorFlow} constructing a graph at the beginning of the execution, whereas \cf\ is based on {\tt pytorch} \cite{NEURIPS2019_9015} and does the batching only with a {\tt Python} {\tt for}-loop with no additional speed-ups. We see that with the largest batch sizes, \cfvII\ fully matches the impressive speed of \cg\ (0.08~ms vs.\ 0.07~ms per shower). 

\begin{figure}[!ht]
  \centering
  \includegraphics[width=0.75\textwidth]{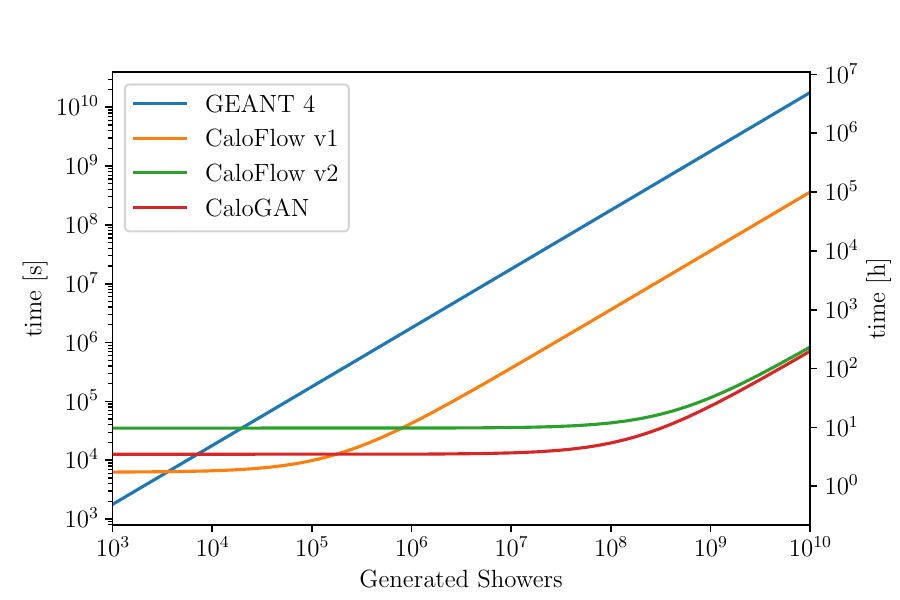}
  \caption{Comparison of shower generation times, using the fastest \cg\ numbers for comparison.}
  \label{fig:timing}
\end{figure}

In fig.~\ref{fig:timing}, we show the time needed to generate the samples vs.\ the size of the requested dataset, including the times needed to train the generative models (visible by the plateau at low number of generated showers).\footnote{Note that fig.~\ref{fig:timing} and tab.~\ref{tab:timing} do not include the time needed to generate the \geant\ training data for the deep generative models.}  Given that many millions (or even billions!) of simulated events are required by the LHC collaborations for their analyses, with each event typically involving hundreds or thousands of showers, this figure demonstrates that the initial computational cost of training the generative models will barely matter when generating samples for actual LHC data analysis. It is clear that fast and accurate \geant\ emulation is an extremely worthwhile endeavor at the LHC~\cite{ATLAS:2021pzo}.

\begin{table}[!t]
  \begin{center}
    \caption{Training and evaluation times of \cf\ and \cg. These are evaluated on a \textsc{Titan V} GPU, the \geant\ runtime is taken from~\cite{Paganini:2017dwg}.}
    \begin{tabular}{|c|c|c|c|c|c|}
      \hline
      & \multicolumn{2}{c|}{\cf} & \multicolumn{2}{c|}{ \cg} & \geant \\
      &\textsc{v1} (teacher) \cite{Krause:2021ilc} &\textsc{v2} (student) &\multicolumn{2}{c|}{  }& \\
      \hline
      \hline
      training & 22+82 min & + 480 min&\multicolumn{2}{c|}{ 210 min } & 0 min \\
      \hline
      \hline
      generation & \multicolumn{5}{c|}{ time per shower} \\
      batch size & &&batch size req. & 100k req. & \\
      \hline
      10    &  835  ms & 5.81 ms & 455  ms& 2.2  ms & 1772 ms\\
      100   &  96.1 ms & 0.60 ms & 45.5 ms& 0.3  ms & 1772 ms\\
      1000  &  41.4 ms & 0.12 ms & 4.6  ms& 0.08 ms & 1772 ms\\
      10000 &  36.2 ms & {\bf 0.08 ms} & 0.5  ms& {\bf 0.07 ms} & 1772 ms\\
      \hline
    \end{tabular}
        \label{tab:timing}
  \end{center}
\end{table}

\section{Conclusions}
\label{sec:conclusions}

In this work, we have presented \cfvII, a faster-sampling normalizing flow for \geant\ calorimeter shower emulation that matches the speed of \cg\ yet retains the superior fidelity of \cfvI\  \cite{Krause:2021ilc}. To achieve this impressive performance, \cfvII\ is based on the fast-sampling IAF architecture, whereas \cfvI\ was based on the alternative MAF architecture. We overcame fundamental obstacles in training IAFs for high dimensional datasets using the novel technique of Probability Density Distillation to fit the ``student'' IAF to the ``teacher'' MAF instead of directly to the \geant\ data. We also improved and innovated beyond the existing ML literature for Probability Density Distillation, inventing several new loss terms that greatly improve the matching of the IAF to the MAF. We expect there could be many applications of this ``fully-guided'' teacher-student training to other domains in fundamental physics and beyond. 

Through \cite{Krause:2021ilc} and the present work, we have demonstrated that normalizing flows are an extremely promising method for fast and accurate generative modeling of high dimensional datasets. With regards to calorimeter emulation, many interesting future directions remain, including generalizing this work to even higher dimensional calorimeters (e.g. ILD~\cite{Buhmann:2020pmy,Buhmann:2021lxj} and CMS HGCAL~\cite{Martelli:2017qbe,Erdmann:2018jxd}), generalizing beyond perpendicular and central incident particles~\cite{deOliveira:2017rwa,Erdmann:2018kuh,Erdmann:2018jxd,ATL-SOFT-PUB-2018-001,Belayneh:2019vyx,ATL-SOFT-PUB-2020-006,ATLAS:2021pzo}, and including simulations of both ECAL and HCAL showers.

\acknowledgments

We are grateful to Ben Nachman for helpful discussions and comments on the draft. This work was supported by DOE grant DOE-SC0010008.  

In this work, we used the {\tt NumPy 1.16.4} \cite{harris2020array}, {\tt Matplotlib 3.1.0} \cite{4160265}, {\tt pandas 0.24.2} \cite{reback2020pandas}, {\tt sklearn 0.21.2} \cite{scikit-learn}, {\tt h5py 2.9.0} \cite{hdf5}, {\tt pytorch 1.7.1} \cite{NEURIPS2019_9015}, and {\tt nflows 0.14} \cite{nflows} software packages. Our code is available at {\tt https://gitlab.com/claudius-krause/caloflow}.

\appendix

\section{Nearest Neighbors}
\label{app:nearest}

\begin{figure}[!ht]
    \centering
    \includegraphics[width=0.75\textwidth, trim= 0 50 25 100, clip]{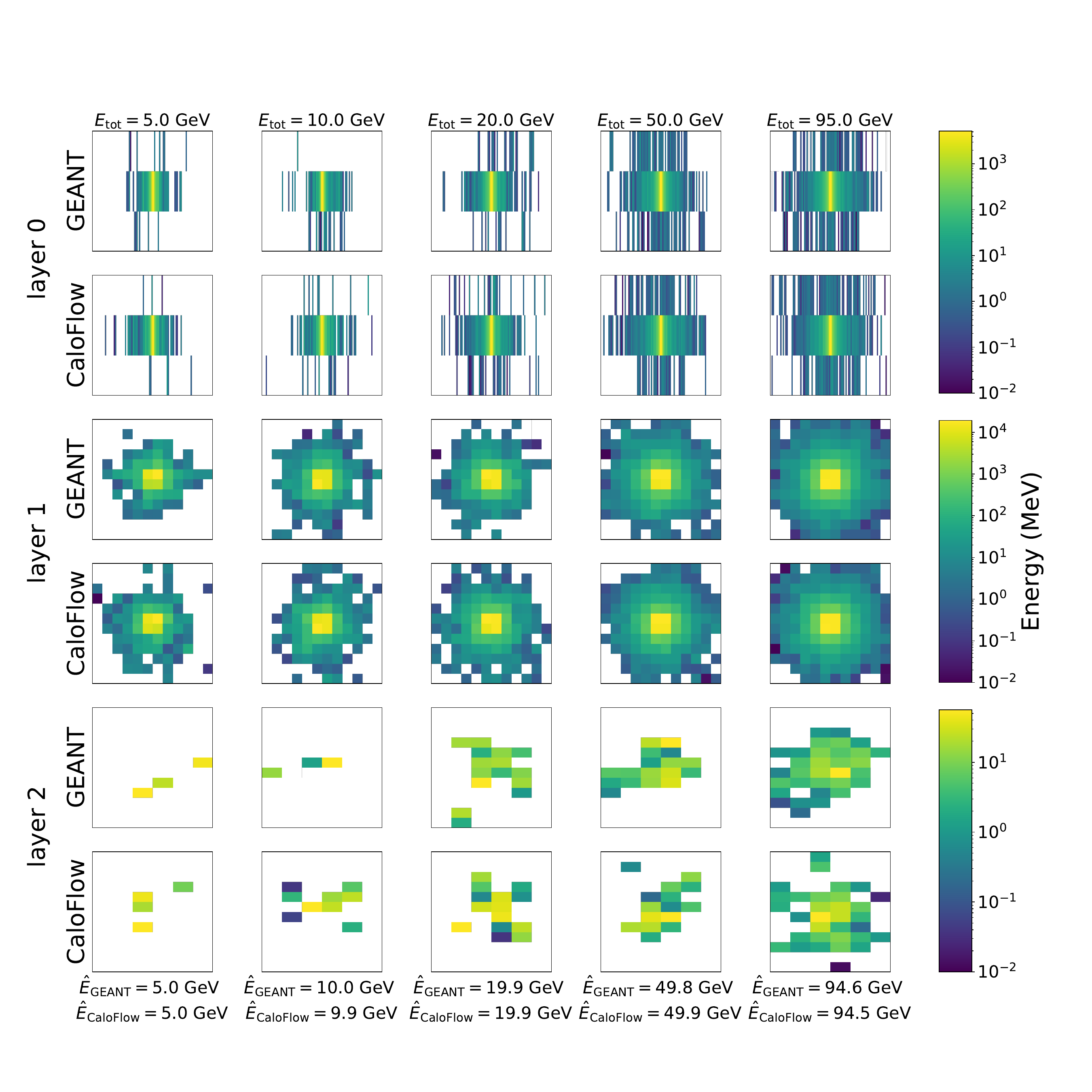}
    \caption{5 randomly selected $e^{+}$ events of \geant\ and their nearest neighbors in the \cf\ student samples. }
    \label{fig:nn.eplus}
  \end{figure}

\begin{figure}[!ht]
    \centering
    \includegraphics[width=0.75\textwidth, trim= 0 50 25 100, clip]{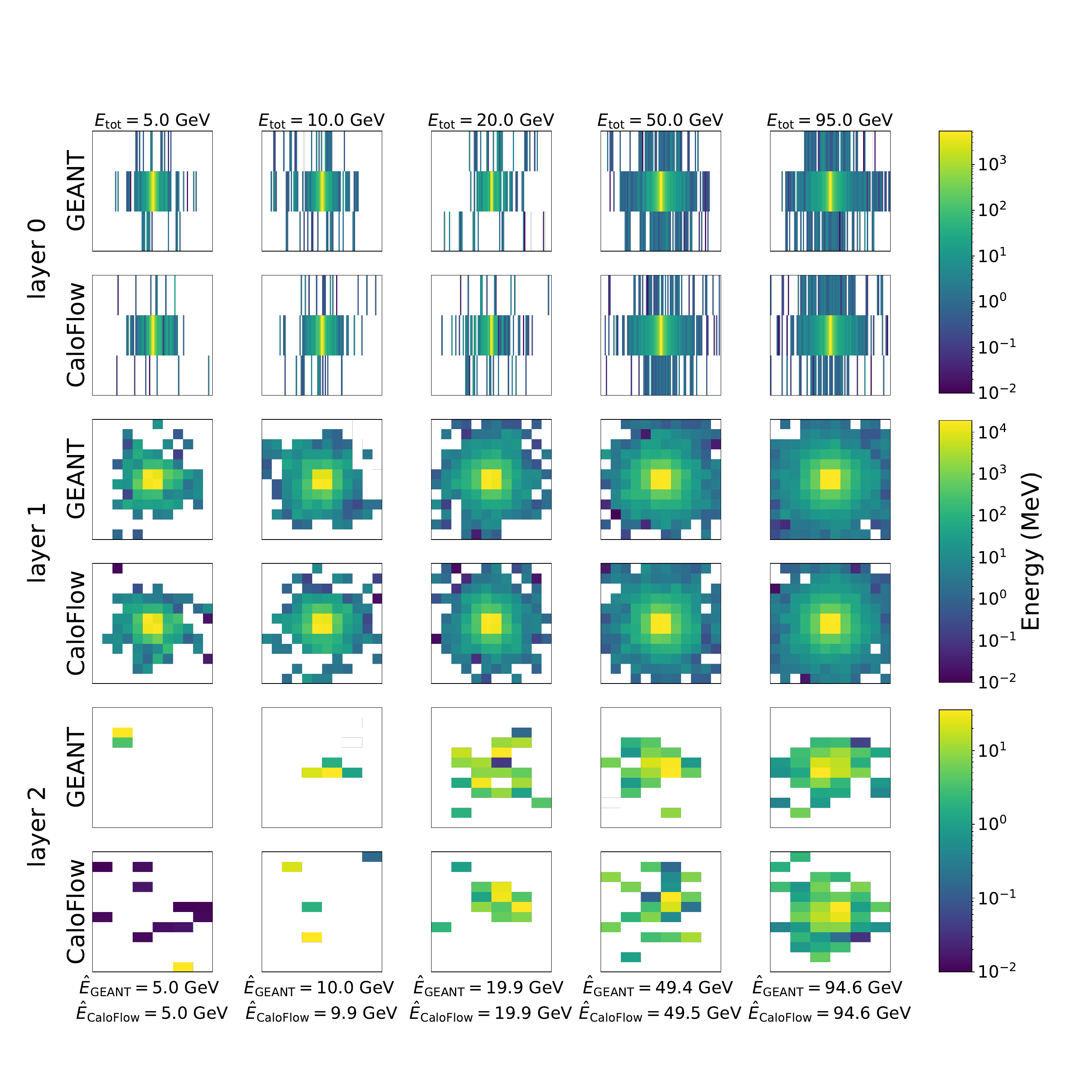}
    \caption{5 randomly selected $\gamma$ events of \geant\ and their nearest neighbors in the \cf\ student samples. }
    \label{fig:nn.gamma}
  \end{figure}

\begin{figure}[!ht]
    \centering
    \includegraphics[width=0.75\textwidth, trim= 0 50 25 100, clip]{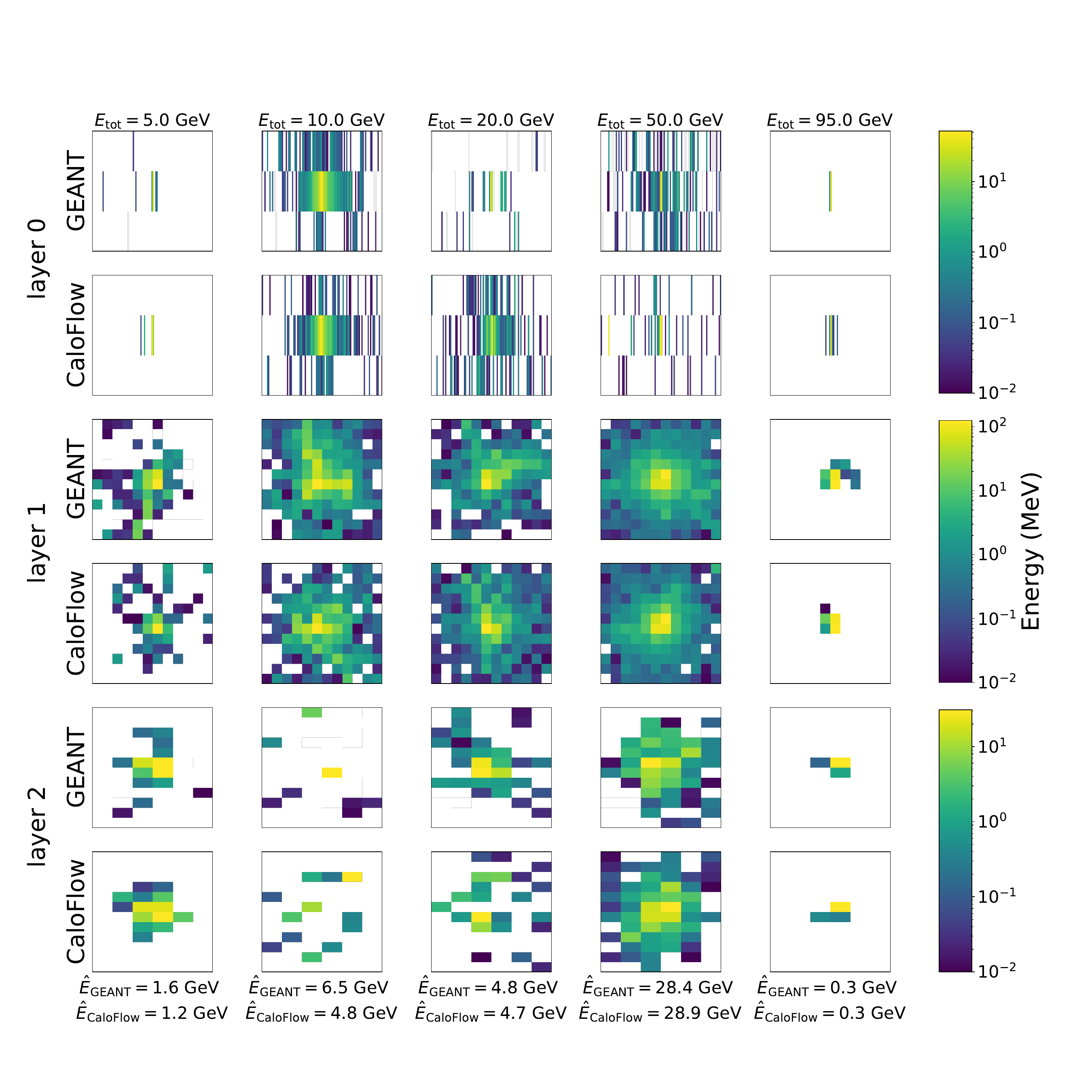}
    \caption{5 randomly selected $\pi^{+}$ events of \geant\ and their nearest neighbors in the \cf\ student samples. }
    \label{fig:nn.piplus}
  \end{figure}

Shown in figs.~\ref{fig:nn.eplus}, \ref{fig:nn.gamma} and \ref{fig:nn.piplus} are 5 randomly selected events from the \geant\ datasets for $e^+$, $\gamma$ and $\pi^+$ at incident energies $E_{\mathrm{inc}} = 5, 10, 20, 50, $ and $95$ GeV and the Euclidean nearest neighbors in the \cf\ student samples. We use the exact same setup as previously in~\cite{Krause:2021ilc}, with 2000 \cf\ samples at each of the incident energies and the exact same \geant\ references. Again, we observe nearest neighbors that are close to the \geant\ samples at all energies, suggesting that no mode collapse occured.

\clearpage

\bibliographystyle{JHEP}
\bibliography{../../calo_lit}

\end{document}